\documentclass[a4paper,onecolumn,preprintnumbers,showpacs,superscriptaddress,aps,prd,11pt]{revtex4}
\usepackage{graphicx}% Include figure files
\usepackage{dcolumn}% Align table columns on decimal point
\usepackage{bm}% bold math
\usepackage{amsfonts,amsmath,amssymb}

\begin{document}
\preprint{AEI-2010-117}
%%%%%%%%%%%%%%%%%%%%%%%%%%%%%%%%%%%%%%%%%%%%%%%%%%%%%%%%%%%%%%%%%%%%%%%%%%%%%%%%
\title{Stable $p$-branes in Chern-Simons AdS supergravities}
%%%%%%%%%%%%%%%%%%%%%%%%%%%%%%%%%%%%%%%%%%%%%%%%%%%%%%%%%%%%%%%%%%%%%%%%%%%%%%%%
\author{Jos\'e D. Edelstein}
\email{jose.edelstein@usc.es}
\affiliation{Department of Particle Physics and IGFAE, University of Santiago de Compostela,
E-15782 Santiago de Compostela, Spain.}
\affiliation{Centro de Estudios Cient\'{\i}ficos (CECS), Casilla 1469, Valdivia, Chile.}
%%%%%%%%%%%%%%%%%%%%%%%%%%%%%%%%%%%%%%%%%%%%%%%%%%%%%%%%%%%%%%%%%%%%%%%%%%%%%%%%
\author{Alan Garbarz}
\email{alan@df.uba.ar}
\affiliation{Departamento de F\'{\i}sica, Universidad de Buenos Aires, Ciudad Universitaria, Pabell\'{o}n 1, 1428, Buenos Aires, Argentina.}
%%%%%%%%%%%%%%%%%%%%%%%%%%%%%%%%%%%%%%%%%%%%%%%%%%%%%%%%%%%%%%%%%%%%%%%%%%%%%%%%
\author{Olivera Mi\v{s}kovi\'{c}}
\email{olivera.miskovic@ucv.cl}
\affiliation{Instituto de F\'{\i}sica, Pontificia Universidad Cat\'{o}lica de Valpara\'{\i}so, Casilla 4059, Valpara\'{\i}so, Chile.}
\affiliation{Max-Planck-Institut f\"{u}r Gravitationsphysik, Albert-Einstein-Institut, Am M\"{u}hlenberg 1, 14476 Golm, Germany.}
%%%%%%%%%%%%%%%%%%%%%%%%%%%%%%%%%%%%%%%%%%%%%%%%%%%%%%%%%%%%%%%%%%%%%%%%%%%%%%%%
\author{Jorge Zanelli}
\email{z@cecs.cl}
\affiliation{Centro de Estudios Cient\'{\i}ficos (CECS), Casilla 1469, Valdivia, Chile.}
%%%%%%%%%%%%%%%%%%%%%%%%%%%%%%%%%%%%%%%%%%%%%%%%%%%%%%%%%%%%%%%%%%%%%%%%%%%%%%%%
\date{\today}

\begin{abstract}
We construct static codimension-two branes in any odd dimension $D$, with negative cosmological constant, and show that they are exact solutions of Chern-Simons (super)gravity theory for (super)AdS$_D$ coupled to external sources. The stability of these solutions is analyzed by counting the number of preserved supersymmetries. It is shown that static massive $(D-3)$-branes are unstable unless some suitable gauge fields are added and the brane is extremal. In particular, in three dimensions, a $0$-brane is recognized as the negative mass counterpart of the BTZ black hole. For these $0$-branes, we write explicitly magnetically charged BPS states with various number of preserved supersymmetries within the $OSp(p|2)\times OSp(q|2)$ supergroups. In five dimensions, we prove that stable $2$-branes with magnetic charge always exist for the generic supergroup $SU(2,2|N)$, where $N \neq 4$. For the special case $N=4$, in which CS supergravity requires the addition of a nontrivial gauge field configuration in order to preserve maximal number of degrees of freedom, we show for two different static 2-branes that they are BPS states (one of which is the ground state), and from the corresponding algebra of charges we show that the energy is bounded from below. In higher dimensions, our results admit a straightforward generalization, although there are presumably more solutions corresponding to different intersections of the elementary objects.
\end{abstract}


\maketitle

\tableofcontents

%%%%%%%%%%%%%%%%%%%%%%%%%%%%%%%%%%%%%%%%%%%%%%%%%%%%%%%%%%%%%%%%%%%%%%%%%%%%%%%%
\section{Introduction}
%%%%%%%%%%%%%%%%%%%%%%%%%%%%%%%%%%%%%%%%%%%%%%%%%%%%%%%%%%%%%%%%%%%%%%%%%%%%%%%%

Strings, membranes and higher dimensional branes in general, are extended objects that generalize the notion of a classical point particle. They are localized objects whose history traces a timelike worldvolume embedded in an ambient spacetime. Although those worldhistories are sets of measure zero, they constitute topological obstruction in spacetime. Their presence leads to a nontrivial classification for the topology of loops and other closed surfaces in spacetime.

These worldhistories are also naked singularities, not surrounded by horizons that prevent an external observer from accessing them. In general relativity, naked singularities are often viewed as unphysical solutions that typically violate some basic laws of physics, from which anything can emerge \cite{Earman1995}. It has been suggested that Nature should prevent the existence of naked singularities through a built-in mechanism inherent to gravitation theory: cosmic censorship.

The simplest example of naked singularity is the static black hole of mass $M<0$. More generally, rotating black holes with angular momentum $J$\ and charge $Q$\ exhibit naked singularities if $M<\sqrt{J^{2}+Q^{2}}$. Even though a general proof of the cosmic censorship conjecture has not been
given, some supporting evidence for it can be presented. It is a simple exercise to show that a static, electrically charged extremal black hole ($|M|=|Q|$) repels a charged particle whose charge ($q$) is larger than its mass ($m$), a mechanism which precludes a charged black hole from becoming
overcharged and naked \cite{Jacobson1996,Crisostomo2004}. Other experiments provide convincing evidence that nonextremal naked singularities are generically unstable under linearized perturbations \cite{Dotti2007}: a small localized perturbation around a naked singularity, grows exponentially in finite time in the linearized approximation for generic initial data.

There is, however, a class of naked singularities that does not give rise to unphysical situations. This was originally discussed in the context of asymptotically AdS$_3$ spacetimes \cite{Izquierdo1995}, where it is well-known that non-singular matter can collapse to a naked singularity \cite{Ross1993,Peleg1995}. The geometry in this case is produced by identifying points in the maximally symmetric manifold $\mathbb{M}$ that are connected by (the exponential action of) a Killing vector $\xi$. The identification involves an operation of cutting and removing a portion of the manifold and sewing it back along identified lines. The resulting quotient space $\widetilde{\mathbb{M}}=\mathbb{M}/\xi$, is generically a new manifold with the same local geometry but different topology. Additionally, if the Killing vector field leaves fixed points, a singularity is introduced where the norm of the Killing vector vanishes, $\xi^{\mu}\,\xi_{\mu }=0$. This vanishing can be interpreted in different ways. If the Killing vector is spacelike, $\xi^{\mu}\,\xi_{\mu }>0$, a fixed point produces a conical singularity; if $\xi^{\mu}\,\xi_{\mu}<0$, a fixed point would generate closed timelike curves, as in the case of the 2+1 black hole, where the region $\xi ^{\mu}\,\xi_{\mu}<0$ corresponds to $r^2<0$ \cite{Banados1992,Banados1993}.

A naked singularity produced by identification in a manifold of constant curvature does not produce a region where the curvature grows infinitely, as in the curvature singularity near $r=0$ in a
four-dimensional Schwarzschild black hole. This means that the conical or causal singularities generated in this form are not infinite sources of energy that could emit unbounded amounts of
energy and paradoxes. Perhaps the most clear example of the physical relevance of conical singularities is in flat 2+1 spacetime, as first shown in \cite{Deser1984}. In fact, a conical singularity cannot be revealed by the local features of the geometry, and only when one takes a vector on a parallel tour around it, something funny happens at the apex of the cone: a finite rotation --equal to the angular deficit that produced the cone. The apex of the cone is not properly a point of the manifold, and its removal changes the topology from $\mathbb{R}^{2}$\ to $\mathbb{R}^{2}\backslash \{0\}$. From the point of view of a physicist living in the vicinity of a conical singularity in an otherwise flat space, it is more useful to describe the conical singularity by saying that the curvature of its manifold has a delta-like singularity, $R_{~2}^{1} = 2\pi\alpha\,\delta(x,y)\,dx\wedge dy$, where the angular deficit is $2\pi \alpha$. The advantage of this notation is that it indicates the rotation angle that a vector picks up by parallel transport around the apex. The disadvantage is that it generates the impression that $r=0$\ is a point of the manifold where the $\delta$ has support. The equivalence of these two points of view was emphasized long ago by J. A. Wheeler, who showed that a multiply-connected spacetime (wormhole) could support a divergence-free electric field which, for a far away observer, would seem to be produced by an electric charge, a phenomenon dubbed by Wheeler as ``charge without charge'' \cite{Wheeler1962}. Similar ideas of topological structures that mimic particle features have been discussed in different geometrical settings \cite{Sorkin1989}, and possibly go back to the middle of the last century \cite{Finkelstein1959}.

The construction outlined above has an essential feature that will be exploited here. Since the identification by a Killing vector does not change the intrinsic geometry of a manifold, the metric properties in $\mathbb{M}$ are the same as in $\widetilde{\mathbb{M}}=\mathbb{M}/\xi $, except at the isolated fixed points of $\xi $. In the cases under study here, $\mathbb{M}$ is the AdS space and $\mathbf{J}^{ab}=x^{a}\partial^{b}-x^{b}\partial ^{a}$ is a spacelike Killing vector in the embedding flat space, that takes the ($x^{a}$-$x^{b}$)-plane onto itself, leaving fixed the co-dimension two space defined by
$x^{a}=x^{b}=0$. This set of fixed points of the Killing vector corresponds to the worldvolume spanned by a $(D-3)$-brane as it evolves in time in the ambient AdS$_{D}$ spacetime. The procedure could be repeated an arbitrary number of times, leaving additional fixed points in the $(x^{a}$\emph{-}$x^{b})$ plane. Each additional deficit produces a cone at a fixed point, with the only restriction that the sum of all angular deficits is bounded in terms of the topology and open/closed nature of the manifold. For instance, in a $2+1$ dimensional Lorentzian manifold admitting an open space-like slicing, it should not exceed $2\pi$ \cite{Izquierdo1995}.

These defect singularities are not very different from an ordinary boundary or membrane, where the manifold is discontinuous. In all these cases, the singularities are topological obstructions and not features revealed by the local geometry. On the other hand, boundaries, branes and topological singularities in general, play another role in physics: these are the places that contain matter sources. Point charges, conductors, strings and matter distributions are usually depicted as submanifolds in space that evolve as submanifolds of spacetime. Their role is to bring in interactions into otherwise free (or self-interacting) classical field theories. In quantum theory the idealized sources are probes to test the response of the theory to perturbations, and therefore furnish a tool to set up a perturbative expansion to extract physically observable information.

There is a class of theories where these ideas turn out to be particularly relevant. These are the so-called Chern-Simons
(super-)gravities, a special class of gauge theories whose dynamics are entirely given in terms of a gauge connection $\mathbf{A}$ \cite{Zanelli2005}. The addition of extended objects in these theories has been considered a decade ago \cite{Mora2000,Mora2001}. However, including interactions of these with the original connection has been tougher than expected. It is worth mentioning that an approach has been considered where the extended $p$-branes are coupled to differential forms and their interaction is given by topological phases produced by their exchange \cite{Nepomechie1985,Wu1988}. There are two problems related to branes in Chern-Simons (CS) theories. The first is how to couple them in a natural way so that the dynamics is still given in terms of the connection $\mathbf{A}$ and some external current. The second problem has to do with the stability of these structures. One way to ensure stability is by making these branes Bogomol'nyi-Prasad-Sommerfield (BPS) states, that is, configurations that admit globally-defined covariantly constant spinors, also called Killing spinors.

In \cite{Edelstein2008}, it was shown that a naive coupling between a CS supergravity theory and a brane does not work as in the standard case. The problem is as follows: in standard supergravity the supersymmetry transformation takes the form $\delta \psi = (d + \omega_{ab}\,\Gamma^{ab} + f_{abc}\,\Gamma ^{abc} + \cdots)\,\epsilon = 0$, where $\omega$ is the spin connection and $f$ is some combination of the RR field strengths, the torsion, etc. In CS gravity theories, instead of $f$ there appear the non-gravitational components of the connection in the super algebra. As a consequence, the projector that is needed to annihilate half the components of $\epsilon$ in the BPS configuration does not form, which in turn means that the branes would systematically break all the supersymmetries. There is a caveat here: it might be that the intricate dynamical structure of CS theories, with different dynamical sectors, could provide an effective theory in some particular sector, with the right field content to reconstruct the right projector. This is a conjecture, however, that would be highly nontrivial to test. An alternative proposal was presented in the same Ref.\cite{Edelstein2008}, as well as in \cite{Miskovic2009}, and it is investigated further in the present paper. The idea is that branes couple to lower-dimensional CS forms for the same connection, but whose components are restricted to live on the brane worldvolume.

A particularly interesting feature of CS theories --and of their supersymmetric extensions-- is their unique constrained form that depends on the spacetime dimension, and the number of supersymmetries considered. The supersymmetry transformations have the same form in all cases, inherited from the usual gauge transformation, $\delta \mathbf{A} = D \boldsymbol{\epsilon}$,
where $D \boldsymbol{\epsilon} = d \boldsymbol{\epsilon} + \left[ \mathbf{A}, \boldsymbol{\epsilon} \right]$ is the covariant derivative of the super algebra (see, {\it e.g.}, \cite{Troncoso1999}). The gravitino $\psi$ belongs to the gauge connection, $\mathbf{A} = \bar\psi\,\mathbf{Q} - \mathbf{\bar{Q}}\,\psi + \cdots$, so that a bosonic brane configuration ($\psi = 0$) is supersymmetric provided
\begin{equation}
\delta \psi = D \boldsymbol{\epsilon} = 0 ~,
\label{Killing spinor eq}
\end{equation}
where the covariant derivative should be projected along the supercharge generators. In the case of AdS supergravities, the relevant contribution to (\ref{Killing spinor eq}) is the spinorial gauge parameter $\epsilon$, $\boldsymbol{\epsilon} = \bar\epsilon\,\mathbf{Q} - \mathbf{\bar{Q}}\,\epsilon + \cdots$ \cite{Zanelli2005}. Then, the stability of these branes is guaranteed if a nontrivial gauge parameter $\epsilon$ is found in the spacetime surrounding the brane such that the previous equation is satisfied, involving, besides the spacetime connection, also the fields that correspond to different interactions required by the closure of the gauge superalgebra (see, e.g., \cite{Aros2002,Aros2007a}).

As shown in \cite{Miskovic2009a}, in order to find a nonvanishing globally defined Killing spinor, it is sufficient to combine a topological defect with another $U(1)$ ``charge'', namely the angular momentum for rotations that leave a spatial plane invariant. The boosts that change the angular velocity of this plane form an Abelian subgroup. Then, by matching the amount of angular deficit and angular momentum the two effects cancel out and a BPS configuration can be obtained by ensuring (\ref{Killing spinor eq}). Those configurations are precisely extremal $0$-branes, with $M=-|J|$. The conclusion reported in this paper is that it is also always possible to stabilize a topological defect with $J=0$, by switching on an appropriate combination of gauge fields. Indeed, we generalize this framework and formally construct static extremal branes of higher codimensions. More concretely, the contents of the paper are the following.

In Section II we briefly review the relevant aspects of CS supergravities and the coupling of extended objects proposed in \cite{Edelstein2008,Miskovic2009}. We deal with the three dimensional case in Section III, where we show that there are no globally defined Killing spinors in a non-trivial configuration unless we have extended supersymmetry and the inclusion of matter. We present the general case based on the $osp(p|2)\times osp(q|2)$ supergravity and construct the spectrum of BPS 0-branes. They include the BTZ black hole and a family of supersymmetric extremal naked singularities. Section IV is devoted to analyzing the case of static 2-branes in five dimensional AdS supergravity. We must remember at this point that this theory has a complicated set of vacua displaying regions with different numbers of degrees of freedom. We consider, thus, the $su(2,2|N)$ theory and find its BPS states in a generic sector of the theory where the number of degrees of freedom is maximal. We show in detail that for $N=4$ these 2-branes saturate a Bogomol'nyi bound. Section V deals with static codimension two branes in AdS in arbitrary higher odd dimensions. We discuss our results and comment on some future avenues of research in Section VI. The article includes a few appendices that attempt to separately address several technical issues in order to ease the reading of its main core while providing all necessary details to make it as self-contained as possible.

%%%%%%%%%%%%%%%%%%%%%%%%%%%%%%%%%%%%%%%%%%%%%%%%%%%%%%%%%%%%%%%%%%%%%%%%%%%%%%%%
\section{Chern-Simons AdS supergravities}
%%%%%%%%%%%%%%%%%%%%%%%%%%%%%%%%%%%%%%%%%%%%%%%%%%%%%%%%%%%%%%%%%%%%%%%%%%%%%%%%

A Chern-Simons (CS) action in $D=2n+1$ dimensions defines a gauge theory for the connection field $\mathbf{A}$ in a Lie algebra $\mathcal{G}$. The dynamic gauge field $\mathbf{A}$ is coupled to an external source $\mathbf{j}_{[2p]}$ in a gauge-invariant way, as described by the action
\begin{equation}
I\!\left[\mathbf{A},\mathbf{j}_{[2p]}\right] = \kappa \int_{\mathbb{M}} \langle \mathbf{C}_{2n+1}(\mathbf{A}) - \mathbf{j}_{[2p]}\wedge \mathbf{C}_{2p+1}(\mathbf{A})\rangle ~.
\label{action}
\end{equation}
Here $\mathbb{M}$ is a ($2n+1$)-dimensional manifold, the level $\kappa$ is a dimensionless quantized coupling constant, the quantities marked in bold take values in $\mathcal{G}$, and $\langle \cdots \rangle $ denotes an invariant symmetric trace in the algebra. The Chern-Simons density $\mathbf{C}_{2n+1}(\mathbf{A})$ is a ($2n+1$)-form that is polynomial in the connection 1-form $\mathbf{A}$ and the curvature 2-form $\mathbf{F} = d \mathbf{A} + \mathbf{A} \wedge \mathbf{A}$, defined through the relation
\begin{equation}
d\langle \mathbf{C}_{2n+1}(A)\rangle = \frac{1}{n+1}\,\langle \mathbf{F} \wedge \cdots \wedge \mathbf{F} \rangle \equiv \frac{1}{n+1}\,\langle \mathbf{F}^{n+1}\rangle ~.
\label{CS density}
\end{equation}
The external current $\mathbf{j}_{[2p]}$ ($0 \leq p < n$) is a covariantly constant $(2n-2p)$-form, $D\mathbf{j}_{[2p]}=0$. Definition (\ref{CS density}) fixes the CS form modulo an exact form $d\Sigma_{2p}$ that corresponds to a boundary term in the action, which we will neglect in this discussion. In quantum theory the boundary terms must be taken into account, because they provide a well-defined action principle, regularizing the action and its conserved charges \cite{Mora2004,Miskovic2006b,Miskovic2007}.

We restrict to spacetimes with negative cosmological constant $\Lambda = - (D-1)(D-2)/2\ell^{2}$, and, consequently, we consider a gauge group that is a supersymmetric extension of anti-de Sitter (AdS) group $SO(D-1,2)$. We denote the generators of the super AdS by
$$
\mathbf{G}_{K}=\left\{\mathbf{J}_{AB};\mathbf{Q}_{\alpha }^{s},\mathbf{\bar{Q}}_{s}^{\alpha },\mathbf{X}^{K}\right\} ~;
$$
they include the AdS generators $\mathbf{J}_{AB} = - \mathbf{J}_{BA}$ ($A,B = 0,\ldots,D$), or pseudo-rotations that leave invariant the metric $\eta_{AB} = \left( -,+,\ldots,+,- \right)$. These are customarily decomposed as $\mathbf{J}_{ab}$ and $\mathbf{P}_{a} := \mathbf{J}_{aD}$, where $a,b = 0,\ldots,D-1$. The supersymmetric generators are $\mathbf{Q}_{\alpha}^{s}$ and $\mathbf{\bar{Q}}_{s}^{\alpha }$, where $\alpha =1,\ldots,2^{[D/2]}$ is a spinorial index and $s = 1,\ldots,\mathcal{N}$ is an internal group index corresponding to R-symmetry. There is a number of additional bosonic generators collectively represented by $\mathbf{X}^{K}$, necessary for the closure of the super AdS algebra. The number of generators $\mathbf{X}$ and the algebra they close changes with the spacetime dimension, and a classification of the super AdS algebras in all dimensions is given in Ref.\cite{Troncoso1999}.

The Lie-algebra valued connection 1-form $\mathbf{A} := A^{K}\,\mathbf{G}_{K}$ can be spanned with respect to these generators in the form
\begin{equation}
\mathbf{A} = \frac{1}{\ell }\,e^{a}\mathbf{P}_{a}+\frac{1}{2}\,\omega ^{ab}\mathbf{J}_{ab}+\left( \bar{\psi}_{\alpha}^{s}\mathbf{Q}_{s}^{\alpha}-\mathbf{\bar{Q}}_{\alpha }^{s}\psi_{s}^{\alpha}\right) +\mathcal{A}\cdot \mathbf{X} ~,
\label{generalA}
\end{equation}
where $e^{a}$ is the vielbein, $\omega ^{ab}$ is the spin-connection and $\psi_{s}^{\alpha }$ are $\mathcal{N}$ gravitini, and we have denoted by $\mathcal{A}_K$ the gauge fields associated to the internal symmetries generated by $\mathbf{X}^{K}$. The associated curvature 2-form, $\mathbf{F} := F^{K}\mathbf{G}_{K}$, in the bosonic sector ($\psi =0$) then reads
\begin{equation}
\mathbf{F} = \frac{1}{2}\,\left( R^{ab}+\frac{1}{\ell ^{2}}\,e^{a}e^{b}\right) \mathbf{J}_{ab}+\frac{1}{\ell }\,T^{a}\mathbf{P}_{a} + \mathcal{F}\cdot \mathbf{X} ~,
\label{generalF}
\end{equation}
with the Riemann curvature $R^{ab} = d \omega^{ab} + \omega^{ac} \wedge \omega_{c}^{\ b}$, the torsion $T^{a} = D e^{a} = d e^{a} + \omega_{~b}^{a} \wedge e^{b}$, and the remaining terms $\mathcal{F} = d \mathcal{A} + \mathcal{A} \wedge \mathcal{A} + \cdots$. The field equations obtained by varying the action (\ref{action}) with respect to $\mathbf{A}$ take the form
\begin{equation}
\left\langle \mathbf{F}^{p} \wedge \left( \mathbf{F}^{n-p}-\mathbf{j}_{[2p]}\right) \mathbf{G}_{K} \right\rangle = 0 ~, \qquad 0 \leq p < n ~,
\label{eom}
\end{equation}
with the invariant tensor of AdS$_{D}$ defined as
\begin{equation}
\left\langle \mathbf{J}_{A_{1}B_{1}} \cdots \mathbf{J}_{A_{n+1}B_{n+1}} \right\rangle = - \frac{1}{2}\, \varepsilon_{A_{1}B_{1}\cdots A_{n+1}B_{n+1}} ~,
\end{equation}
where we use the convention $\varepsilon ^{01\cdots D}=1$. The locally flat configuration, $\mathbf{F}=0$ (pure gauge), is always a solution of the source-free CS equation $\left\langle \mathbf{F}^{n}\,\mathbf{G}_{K}\right\rangle =0$, while in presence of the sources the pure gauge becomes solution only for $p\neq 0$.

We consider a particular form of the current $\mathbf{j}_{[2p]}$ that describes a $2p$-brane with charge $q_{2p}$, localized on a ($2p+1$)-dimensional time-like manifold $\Gamma_{2p+1}$ representing its worldvolume. Being charged with respect to the gauge group implies that this current transforms in some nontrivial representation of the Lie algebra, labeled by a set of $n-p$ indices,
\begin{equation}
\mathbf{j}_{[2p]} = q_{2p}\,\delta(\Sigma_{2n-2p})\,\mathbf{G}^{K_{1} \cdots
K_{n-p}} ~.
\label{generic j}
\end{equation}
The Dirac delta in this equation is a ($2n-2p$)-form with support at the center of the $(2n-2p)$-dimensional transverse spacelike manifold $\Sigma_{2n-2p}$. The particular form of the current $\mathbf{j}_{[2p]}$ given by Eq.(\ref{generic j}) represents a static $2p$-brane at the center of the transverse space.

The interaction with the brane breaks the gauge symmetry. The worldvolume of the brane could have at most local $SO(2p,2)$\ isometries, while the transverse section is at most invariant under $SO(2n-2p)$. Thus, the interaction reduces the maximal $SO(2n,2)$ isometry of $\mathbb{M}^{2n+1}$ down to a maximal isometry $SO(2p,2) \times SO(2n-2p)$ of the $2p$-brane spacetime $\Gamma_{2p+1}\times \Sigma_{2n-2p}$. For supersymmetric solutions, we expect that the fermionic part of the gauge symmetry group behaves similarly, being broken by the very presence of a static $2p$-brane to a supergroup whose bosonic part is $SO(2p,2) \times SO(2n-2p)$.

%%%%%%%%%%%%%%%%%%%%%%%%%%%%%%%%%%%%%%%%%%%%%%%%%%%%%%%%%%%%%%%%%%%%%%%%%%%%%%%%
\section{Static 0-brane in AdS$_{3}$}
%%%%%%%%%%%%%%%%%%%%%%%%%%%%%%%%%%%%%%%%%%%%%%%%%%%%%%%%%%%%%%%%%%%%%%%%%%%%%%%%

We deal now with the case of three dimensional AdS supergravities. Even if much is known about them, we review some of these results, identifying new supersymmetric extremal naked singularities that will be interpreted as BPS 0-branes, whose spectra we determine.

%%%%%%%%%%%%%%%%%%%%%%%%%%%%%%%%%%%%%%%%%%%%%%%%%%%%%%%%%%%%%%%%%%%%%%%%%%%%%%%%
\subsection{Conical defect in AdS$_{3}$ (Review)}
%%%%%%%%%%%%%%%%%%%%%%%%%%%%%%%%%%%%%%%%%%%%%%%%%%%%%%%%%%%%%%%%%%%%%%%%%%%%%%%%

AdS$_{3}$ space can be seen as a hyperboloid in $\mathbb{R}^{2,2}$, where the Cartesian coordinates $x^{A}=(x^{0},x^{1},x^{2},x^{3})$ are subjected to the constraint $\eta_{AB}\,x^{A}x^{B}=-\ell ^{2}$. Using the parametrization
\begin{equation}
\begin{array}{ll}
x^{0} = A \cos \phi_{03} ~, \qquad\qquad & x^{1} = B \cos \phi_{12} ~, \\ [0.8em]
x^{3} = A \sin \phi_{03} ~, \qquad\qquad & x^{2} = B \sin \phi_{12} ~,
\end{array}
\label{3dstatic}
\end{equation}
the AdS$_{3}$ space corresponds to the surface $B^{2} - A^{2} = -\ell^{2}$. Its metric then reads
\begin{equation}
ds^{2} = \eta_{AB}\,dx^{A}\,dx^{B} = \frac{\ell ^{2}}{B^{2}+\ell ^{2}}\,dB^{2}-\left( B^{2}+\ell^{2}\right) \,d\phi_{03}^{2} + B^{2} d\phi_{12}^{2} ~.
\label{AdS metric}
\end{equation}
Note that by unwrapping the $\phi_{03}$ coordinate and calling $B=r$, one gets a global covering of AdS$_{3}$ in polar coordinates.

A 0-brane can be seen as a defect on the $(x^{1}$-$x^{2})$-plane produced by an angular deficit of $2\pi \alpha$ in the $\phi_{12}$ angle, so that
\begin{equation}
\phi_{12} \simeq \phi _{12} + 2 \pi (1 - \alpha) ~.
\label{defect}
\end{equation}
A natural way to implement this is by introducing a scaled coordinate $\phi$ such that $\phi_{12} = (1-\alpha)\,\phi $, where $\phi \simeq \phi +2\pi $. Introducing the rescaled radial and time coordinates, $r = (1-\alpha)\,B$ and
$t = \ell\,\phi_{03}/(1-\alpha)$, respectively, the metric (\ref{AdS metric}) becomes
\begin{equation}
ds^{2} = -\left( (1-\alpha)^{2}+\frac{r^{2}}{\ell^{2}}\right) dt^{2}+\frac{dr^2}{(1-\alpha)^{2} + \frac{r^{2}}{\ell^{2}}} + r^{2} d\phi^{2} ~,
\label{ds2}
\end{equation}
which is just the Ba\~nados-Teitelboim-Zanelli (BTZ) solution, but with negative mass, $M = -(1 - \alpha)^{2}$. Therefore, this naked singularity 0-brane sits at the topological defect whose magnitude $2 \pi \alpha$ is related to the ``negative mass of the black hole'' \cite{Izquierdo1995,Miskovic2009a}.

The identification (\ref{defect}) in terms of the coordinates of the embedding space represents a $\phi_{12}$-rotation by $2 \pi (1-\alpha)$ in the $(x^{1} $-$x^{2})$ plane,
\begin{equation}
\left( \begin{array}{c} x^{1} \\ x^{2} \end{array} \right) \simeq \left( \begin{array}{cc} \cos 2\pi \alpha & \sin 2\pi \alpha \\ - \sin 2\pi \alpha & \cos 2\pi \alpha \end{array} \right) \left( \begin{array}{c} x^{1} \\ x^{2} \end{array} \right) = e^{-2\pi \alpha \,\mathbf{J}_{12}}\, \left( \begin{array}{c} x^{1} \\ x^{2} \end{array} \right) ~,
\label{Xc,Yc}
\end{equation}
where $\mathbf{J}_{12} = \left( \begin{array}{cc} 0 & -1 \\ 1 & 0 \end{array} \right)$ is a matrix representation of $\partial_{\phi }$ acting on the vector $(x^{1},x^{2})$, or in its more convenient unitary representation, $\mathbf{J}_{12} = x^{1}\,\partial_{2} - x^{2}\,\partial_{1}$. The exponent is the Killing vector $\xi =\xi ^{A}\partial _{A}$ that identifies $x^{A}$ and $x^{A}+\xi ^{A}$,
\begin{equation}
\xi =2\pi \alpha \,(x^{2}\partial _{1}-x^{1}\partial _{2})=-2\pi \alpha\,\partial _{\phi} ~.
\label{identification}
\end{equation}

Note that for $\alpha = 0$ (no deficit) the Killing vector vanishes. Also note that there is an ambiguity in the choice of $\alpha$: shifting $\alpha$ by an integer produces the same effect in (\ref{Xc,Yc}). This means that, for an identification, $\alpha$ and $\alpha + k$ are indistinguishable. Nevertheless, a deficit angle greater than $2\pi$ seems geometrically inadmissible and therefore we consider $\alpha $ to be restricted to values $\leq 1$.

On the other hand, there is no such restriction for an angular excess, that corresponds to a negative value of $\alpha$ in the metric, or in Eq.(\ref{defect}). Therefore, the identity $2\pi \alpha \simeq 2\pi (\alpha +k)$ seems to be acceptable for $\alpha \in \lbrack 0,1]$ and negative integer values of $k$. This does not correspond to a standard conical defect but rather to a {\em lettuce leaf}-like configuration known as Elizabethian geometry \cite{Muller2008}.

There is another discrete symmetry related to the choice of $\alpha$ in the metric: reflection $(1-\alpha)\rightarrow -(1-\alpha)$ is equivalent to the shift $\alpha \rightarrow 2-\alpha$, that cannot be obtained from $\alpha$ by addition of any integer $k$. The curvature is invariant under this symmetry --for example, a $D$-dimensional 0-brane obtained as a surface deficit of $S^{D-2}$  has the scalar curvature
$$
R = \Lambda -\frac{\left( D-2 \right) \left( D-3\right) \alpha \left( \alpha -2\right) }{\left( 1-\alpha\right) ^{2} r^2} ~,
$$
that clearly becomes the constant $\Lambda$ when $D=3$, or when there is no deficit $\alpha =0$, but also when $\alpha =2$.

In what follows, we will discuss only the values of $\alpha$ that correspond to angular deficits, namely, $\alpha \in [0,1)$. The spectrum of three-dimensional naked singularities is the one described originally in \cite{Izquierdo1995}.

%%%%%%%%%%%%%%%%%%%%%%%%%%%%%%%%%%%%%%%%%%%%%%%%%%%%%%%%%%%%%%%%%%%%%%%%%%%%%%%%
\subsection{Field equations}
%%%%%%%%%%%%%%%%%%%%%%%%%%%%%%%%%%%%%%%%%%%%%%%%%%%%%%%%%%%%%%%%%%%%%%%%%%%%%%%%

A straightforward computation shows that the geometry defined by (\ref{ds2}) is a spacetime that is AdS almost everywhere and torsion free,
\begin{eqnarray}
& & R^{ab}+\frac{1}{\ell^{2}}\,e^{a}\wedge e^{b} = 2 \pi \alpha\,\delta(\Sigma_{12})\,dx^{1} \wedge dx^{2}\,\delta_{[12]}^{[ab]} ~, \quad \\ [1em]
& & T^{a} = 0 ~,
\end{eqnarray}
where the Dirac delta has support at the center of the two-dimensional spatial section that corresponds to the $(x^{1}$-$x^{2})$-plane. These two equations can be put in a more compact form as
\begin{equation}
\mathbf{F}=\mathbf{j}_{[0]} ~,
\label{F=J}
\end{equation}
where the source $\mathbf{j}_{[0]}$ is a Lie-algebra valued 2-form current
\begin{equation}
\mathbf{j}_{[0]} = 2 \pi \alpha\,\delta(\Sigma_{12})\,\mathbf{J}_{12} ~,
\label{AdS 3current}
\end{equation}
that indicates the presence of a brane at the center of the $\Sigma_{12}$, produced by a deficit angle $2\pi\alpha $ generated by $\mathbf{J}_{12}$ (see Appendix \ref{RegCone} for further details). This is the field equation that governs the spacetime geometry in the presence of a source, that can be obtained from the general expression for the field equations, Eq.(\ref{eom}), in the only possible case in three dimensions $n=1$, $p=0$.

The stability of a 0-brane can be established if the background can be viewed as the bosonic sector of a supersymmetric configuration (BPS state). For this, it is sufficient to prove the existence of a globally defined Killing spinor field that satisfies (\ref{Killing spinor eq}) in the spacetime that surrounds the brane. The worldvolume of the brane is not part of the ambient manifold $\mathbb{M}$ because the geometry is not properly defined there. The situation is analogous to trying to solve Dirac equation on a conical surface: clearly one
does not worry about the behavior of the Dirac field \emph{at} the conical singularity, since that point is not properly part of a differentiable manifold.

In order to inspect the BPS condition, one needs to know the explicit form of the gauge algebra that extends the AdS symmetry in the corresponding dimension.

%%%%%%%%%%%%%%%%%%%%%%%%%%%%%%%%%%%%%%%%%%%%%%%%%%%%%%%%%%%%%%%%%%%%%%%%%%%%%%%%
\subsection{BPS branes in 2+1 dimensions}
%%%%%%%%%%%%%%%%%%%%%%%%%%%%%%%%%%%%%%%%%%%%%%%%%%%%%%%%%%%%%%%%%%%%%%%%%%%%%%%%

Supersymmetric extension of AdS group in three dimensions, with $\mathcal{N} = p+q$ supersymmetries, is $OSp(p|2)\times OSp(q|2)$ \cite{Achucarro1989a} with the generators $\mathbf{G}_{K} = \left\{ \mathbf{G}_{K}^{+},\mathbf{G}_{K}^{-} \right\}$, and the connection 1-form can be written as
\begin{equation}
\mathbf{A}=A^{K}\mathbf{G}_{K}=\mathbf{A}^{+}+\mathbf{A}^{-} ~,
\end{equation}
where
\begin{equation}
\mathbf{A}^{\pm }=\left( \omega ^{\,a}\pm \frac{1}{\ell }\,e^{\,a}\right) \mathbf{J}_{a}^{\pm }+\frac{1}{2}\,b_{\pm }^{IJ}\,\mathbf{T}_{IJ}^{\pm}+\psi_{\pm \alpha }^{I}\,\mathbf{Q}_{I}^{\pm \alpha} ~.
\end{equation}
Here $\{\mathbf{T}_{IJ}^{+},\mathbf{T}_{I^{\prime }J^{\prime }}^{-}\}$ generate the\textbf{\ }$O(p)\times O(q)$ subgroup, and we have introduced $\omega ^{a}=\frac{1}{2}\,\varepsilon _{abc}\omega ^{bc}$. The corresponding field strength also splits as $\mathbf{F}=\mathbf{F}^{+}+\mathbf{F}^{-}$,
with
\begin{equation}
\mathbf{F}^{\pm} = \displaystyle\left( R^{a} \pm \frac{1}{\ell}\,T^{a} + \frac{1}{2\ell^{2}}\,\varepsilon _{\ \ bc}^{a}\,e^{b}\wedge e^{c}\right) \mathbf{J}_{a}^{\pm } + \frac{1}{2}\,\mathcal{F}_{\pm }^{IJ}\,\mathbf{T}_{IJ}^{\pm} + \text{spinors} ~,
\end{equation}
the curvature being given by $R^{a}=\frac{1}{2}\,\varepsilon_{abc}\,R^{bc} = d \omega ^{a} + \frac{1}{2} \varepsilon ^{abc} \omega_{b} \wedge \omega_{c} $ (see Appendix \ref{Super AdS3} for details).

We seek for a bosonic configuration ($\psi^{\pm} = 0$) that possesses nontrivial supersymmetries $\boldsymbol{\epsilon} = \boldsymbol{\epsilon}^{+} + \boldsymbol{\epsilon}^{-} = \epsilon_{I}^{+\alpha}\,\mathbf{Q}_{\alpha}^{+I} + \epsilon_{I^{\prime}}^{-\alpha}\,\mathbf{Q}_{\alpha}^{-I^{\prime}}$, so that the spinor $\boldsymbol{\epsilon}$ is a solution of the Killing spinor equation (\ref{Killing spinor eq}),
$$
D\boldsymbol{\epsilon} := \left( D_{+} \epsilon^+\right)
_{I}^{\ \alpha} \,\mathbf{Q}_{\alpha }^{+I} + \left(
D_{-} \epsilon^-\right)_{I^{\prime}}^{\ \alpha}\,\mathbf{Q}_{\alpha
}^{-I^{\prime }} = 0 ~.
$$
Each term must be zero independently, so we have
\begin{equation}
D_{\pm} \epsilon^{\pm }=\left[ d-\frac{1}{2}(\omega ^{a}\pm
\frac{1}{\ell }\,e^{a})\Gamma_{a} + b_{^{\pm}} \right]
\epsilon^{\pm} = 0 ~,
\label{Killingpm}
\end{equation}
where $b_{^{\pm}}$ is a square matrix with components $(b_{^{\pm}})_{\ L}^{K} $ (the components of the $o(p)$ or $o(q)$ gauge fields), and $\Gamma_{a}$ are Dirac matrices. The term $b_{^\pm}\epsilon^\pm$ means $(b_{^{\pm}}\epsilon^\pm)_{\ \alpha}^{I}=(b_{^\pm})^I_{\ J}\epsilon^{\pm J}_{\ \alpha}$.

The AdS connection in the region around the brane is locally flat, $\mathbf{F}^{\pm} = 0$. This means that the torsion must vanish and the metric is that of a locally AdS spacetime. The only effect of the presence of the brane is in the topology of the region around it. Next, the conditions for the geometry to admit a global Killing spinor in the presence of the defect will be analyzed.

%%%%%%%%%%%%%%%%%%%%%%%%%%%%%%%%%%%%%%%%%%%%%%%%%%%%%%%%%%%%%%%%%%%%%%%%%%%%%%%%
\subsubsection{ $\mathcal{N}=1$ supersymmetry}
%%%%%%%%%%%%%%%%%%%%%%%%%%%%%%%%%%%%%%%%%%%%%%%%%%%%%%%%%%%%%%%%%%%%%%%%%%%%%%%%

The minimal supersymmetry, $\mathcal{N}=1$, is described by the super AdS algebra $osp(1|2)$ with $(p,q) = (1,0)$ or $(0,1)$, so the $b^{\pm}$ are absent, and only one gravitino, either $\psi ^{+}$ or $\psi ^{-}$, is present, and consequently, either $\mathbf{Q}^{+}$ or $\mathbf{Q}^{-}$ is included. Suppose the supersymmetry generator is $\mathbf{Q}^{+}$; then, the Killing spinor is $\boldsymbol{\epsilon} = \epsilon^{+}\,\mathbf{Q}^{+}$ and equation (\ref{Killingpm}) must be solved for the $+$ choice only. The derivation is presented in Appendix \ref{SolKilling3D}, and the result is given by
\begin{equation}
\epsilon^{+} = e^{f(r)\Gamma_{1}}\, e^{\frac{1}{2}\,i(1-\alpha)\left( \phi +\frac{t}{\ell} \right)}\,\eta ^{+} ~,
\label{psi(1,0)}
\end{equation}
where $(t,r,\phi) = \left( \frac{\ell \phi_{03}}{1-\alpha},(1-\alpha)\,B,\frac{\phi_{12}}{1-\alpha} \right)$, $f(r)=\frac{1}{2}\,\sinh^{-1}\!\!\left( \frac{r}{(1-\alpha)\ell} \right)$, and $\eta ^{+}$ is a constant eigen-spinor of $\Gamma_{0}$ (say, $\Gamma_{0}\,\eta ^{+}=i\eta ^{+}$). Similarly, for the $(0,1)$ spinor $\boldsymbol{\epsilon} = \epsilon^{-}\,\mathbf{Q}^{-}$, one obtains
\begin{equation}
\epsilon^{-}=e^{-f(r)\Gamma _{1}}e^{-\frac{1}{2}\,i(1-\alpha)\left( \phi +\frac{t}{\ell}\right) }\eta ^{-} ~,
\label{psi(0,1)}
\end{equation}
where $\Gamma_{0}\,\eta^{-} = - i \eta ^{-}$ (of course, one could choose the $+i$ eigenvalue as well). In both cases, the spinor $\epsilon$ must be either periodic or antiperiodic in the angular coordinate $\phi$, $\epsilon (\phi + 2 \pi) = \pm \epsilon (\phi)$. The expressions (\ref{psi(1,0)}) and (\ref{psi(0,1)}) satisfy these boundary conditions provided $a$ is an integer and therefore,
\begin{equation}
\alpha = n \in \mathbb{Z} ~.
\end{equation}
Since the angular deficit satisfies $\alpha \in \lbrack 0,1]$, the only BPS configurations are either $\alpha =0$ (spacetime is AdS$_{3}$, no defect), or $\alpha =1$ (the spacetime is the zero-mass BTZ black hole). This means that there are no globally defined Killing spinors except in the known cases ($M=0,-1$), as reported in \cite{Coussaert1994}.

%%%%%%%%%%%%%%%%%%%%%%%%%%%%%%%%%%%%%%%%%%%%%%%%%%%%%%%%%%%%%%%%%%%%%%%%%%%%%%%%
\subsubsection{$\mathcal{N}=2$ supersymmetries}
%%%%%%%%%%%%%%%%%%%%%%%%%%%%%%%%%%%%%%%%%%%%%%%%%%%%%%%%%%%%%%%%%%%%%%%%%%%%%%%%

$\mathcal{N}=2$ supersymmetries occur for $(p,q) = (1,1)$, $(2,0)$ and its symmetric reflection, $(0,2)$. The case $(p,q) = (1,1)$ admits the Killing spinor $\boldsymbol{\epsilon} = \epsilon ^{+}\,\mathbf{Q}^{+} + \epsilon^{-}\,\mathbf{Q}^{-}$, where $\epsilon^{+}$ and $\epsilon^{-}$ are given by Eqs. (\ref{psi(1,0)}) and (\ref{psi(0,1)}), respectively, which again implies either $\alpha =0$ or $1$.

In the case $(p,q) = (2,0)$, the algebra contains a generator of
$o(2)$ that (modulo reflections) acts as $u(1)$. The corresponding
Abelian field, $\mathbf{b}$, introduces an additional charge in one
of the two copies (say, $\epsilon^{+}$). In this representation,
$(b_{+})^{I}_{\ J}:=- b\,\sigma _{\ J}^{I}$, where $\sigma =\left(
\begin{array}{cc} 0 & 1 \\ -1 & 0
\end{array}\right)$. The CS field equations around the source are
$\mathbf{F}^{\pm }=0$, where the curvatures read
\begin{eqnarray}
\left( \mathbf{F}^{+}\right)_{I}^{\ J} & = & \delta_{I}^{\ J} \left( R^{a} + \frac{1}{\ell }\,T^{a} +
\frac{1}{2\ell^{2}}\,\varepsilon_{\,\,\,bc}^{a} e^{b} \wedge e^{c}
\right) \mathbf{J}_{a}^{+} - \frac{1}{2}\,db\,\sigma_{I}^{\ J} ~, \\ [0.8em]
\left( \mathbf{F}^{-}\right)_{I}^{\ J} & = & \delta _{I}^{\ J}\! \left(
R^{a} - \frac{1}{\ell}\,T^{a} +
\frac{1}{2\ell^{2}}\,\varepsilon_{\,\,\,bc}^{a} e^{b} \wedge
e^{c}\right) \mathbf{J}_{a}^{-} ~.
\end{eqnarray}
Therefore, the geometry is locally AdS and torsion-free as in the previous case, and $db=0$. The last condition enables us to write the $1$-form $b$ locally as $b = d\Omega$. Globally, this is much more interesting than being a trivial connection, since $\Omega$ could be multivalued (like the angle $\phi_{12}$ itself), allowing for different topological sectors for $b$, labelled by the winding number. This provides the basics to find a non-trivial Killing spinor charged with respect to $b$, producing a Bohm-Aharonov phase that cancels the contribution of the spin connection \cite{Edelstein1996,Edelstein1996b}. Thus, a Killing spinor $\boldsymbol{\epsilon} = \epsilon_{I}^{+}\,\mathbf{Q}^{+I}$
satisfies
\begin{equation}
d \epsilon_{I}^{+} - \frac{1}{2} \left( \omega ^{a} + \frac{1}{\ell}\,e^{a} \right) \Gamma_{a}\,\epsilon_{I}^{+} - d\Omega\ \sigma^{\ J}_{I}\,\epsilon_{J}^{+} = 0 ~.
\label{DE+}
\end{equation}
Choosing $\Omega = q\,\phi_{12}$, only one component of the Killing equation receives a correction when compared with its form for $b=0$,
\begin{equation}
\left( \partial_{\phi_{12}} - \frac{1}{2}\,\Gamma_{0} - \,q\,\sigma
\right) \lambda(\phi_{12}) = 0 ~.
\label{dLambda}
\end{equation}
The solution is
\begin{equation}
\epsilon_{I}^{+} = e^{f(r)}\ e^{\frac{i}{2\ell}\,(1-\alpha) t\,+\frac{i}{2}\,(1-\alpha) (1 + 2 q)\phi}\ \eta_{I}^{+} ~,
\label{psi(2,0)}
\end{equation}
where we have used $\phi_{12}=(1-\alpha)\phi$ and that $\eta_{I}^{+}$ is a constant simultaneous eigenspintor of $\sigma$ and $\Gamma_{0}$,
\begin{equation}
\sigma^{\ J}_{I}\,\eta_{J}^{+} = i \eta_{I}^{+} ~, \qquad \left( \Gamma_{0}\right)_{\ \beta}^{\alpha} (\eta_{I}^{+})^{\beta} = i (\eta_{I}^{+})^{\alpha} ~.
\label{double eigen}
\end{equation}
The (anti-)periodic boundary condition $\epsilon^{+}(\phi + 2 \pi) = \pm \epsilon^{+}(\phi)$ requires the $U(1)$ charge to be quantized,
\begin{equation}
(1-\alpha)\,(1+2q) \in \mathbb{Z} ~.
\label{charge quantization}
\end{equation}
Notice that this extremality condition perfectly matches that obtained by Izquierdo and Townsend (after replacing $(1 - \alpha) \to \beta$ and $(1 - \alpha)\, q \to Q$ in their eq.(3.5)) \cite{Izquierdo1995}. Therefore, for a given topological defect $\alpha \in [0,1)$, all charges given by $q = \frac{k}{2(1-\alpha)} - \frac{1}{2}$, $k \in \mathbb{Z}$ satisfy the BPS condition. Conversely, if the $U(1)$ charge is fixed, there are several possible values for angular defect given by
\begin{equation}
0 < \alpha_{k} = \frac{2q+k}{2q+1} < 1 ~, \qquad k \in \mathbb{Z} ~.
\label{alfarange}
\end{equation}
Note that for a given value of $q$, the number of allowed values for $\alpha$ increase with $|q|$.

We conclude that nontrivial Killing spinors exist for these choices of $q$ and $\alpha $, and the corresponding $0$-branes should be stable BPS configurations. Each matrix condition in (\ref{double eigen}) projects out $1/2$ of the spinor components, so the final solution preserves $1/4$ of the original supersymmetries; a $1/4$-BPS state. There is a single unbroken supercharge in the solution. Obviously, the same is true for $(p,q) = (0,2)$, just replacing $+$ by $-$ in the preceding discussion.

The current that describes this $0$-brane couples to the geometry and to the $U(1)$ field. The gravitational part of the current is given by Eq.(\ref{AdS 3current}). Additionally, the $U(1)$ charge of the brane couples to $b$. The form of this contribution can be found from the Abelian gauge field, $b = q\,d\phi_{12}$ (that carries an electromagnetic flux given by the integral of $q\, dd\phi_{12} = - 2 \pi \alpha\,q\,\delta(\Sigma_{12})$), so the total current is (see (\ref{F=J}))
\begin{equation}
\mathbf{j}_{[0]} = 2\pi \alpha \left( \mathbf{J}_{12}-q\mathbf{T}_{+}^{12}\right)\,\delta(\Sigma_{12}) ~.
\end{equation}
The presence of the $o(2)$ $R$-symmetry field $b$ is responsible for stabilizing the 0-brane: the conical defect in the spatial section is compensated by the $U(1)$ charge in the internal gauge space \cite{Edelstein1996,Edelstein1996b}. In the following sections, it is shown that this is a generic feature, and that an Abelian gauge field can stabilize any static codimension-two brane in higher dimensions.

%%%%%%%%%%%%%%%%%%%%%%%%%%%%%%%%%%%%%%%%%%%%%%%%%%%%%%%%%%%%%%%%%%%%%%%%%%%%%%%%
\subsubsection{$\mathcal{N}=p+q$ supersymmetries}
%%%%%%%%%%%%%%%%%%%%%%%%%%%%%%%%%%%%%%%%%%%%%%%%%%%%%%%%%%%%%%%%%%%%%%%%%%%%%%%%

For the $osp(p|2)\times osp(q|2)$ superalgebra, the brane solution is again locally flat, $\mathbf{F} = 0$, namely, locally AdS geometry, torsion-free, and a flat $R$-connection $db_{(\pm )J}^{I} + b_{(\pm )K}^{I} \wedge b_{(\pm)J}^{K} = 0$,
\begin{equation}
\mathbf{A}_{\text{AdS}}=0\text{-brane} ~, \qquad \,b_{\pm}^{IJ}: \text{locally flat} ~,
\end{equation}
where the 0-brane is given by Eqs. (\ref{AdS metric}, \ref{defect}). The connection $\mathbf{b}$ has the general form $\mathbf{b=g}^{-1} d\mathbf{g}$ (where $\mathbf{g}$ belongs to $O(p)\times O(q)$), but here we consider a particular Abelian choice of this form in the Cartan subalgebra of $o(p)\times o(q)$, such that $db_{(\pm)J}^{I}=0$, and $b_{(\pm)K}^{I} \wedge b_{(\pm)J}^{K} = 0$. The
Cartan subalgebra is spanned by
$$
\left\{ \mathbf{T}_{12}^{+}, \mathbf{T}_{34}^{+}, \ldots, \mathbf{T}_{2[\frac{p}{2}]-1,2[\frac{p}{2}]}^{+}; \mathbf{T}_{12}^{-}, \mathbf{T}_{34}^{-}, \ldots ,\mathbf{T}_{2[\frac{q}{2}]-1,2[\frac{q}{2}]}^{-} \right\} ~.
$$
This means that we can take the matter connection as $\mathbf{b} = - \mathbf{T}\,d\phi_{12}$, with $\mathbf{T}$ a linear combination of some Cartan generators, say $k_{+} + k_{-}$ of them, with $k_+ \leq [p/2]$ and $k_- \leq [q/2]$, and the coefficients represent the corresponding charges $q_k^{\pm}$. Explicitly,
\begin{equation}
\mathbf{T} = \mathbf{T}_{+} + \mathbf{T}_{-} ~,
\end{equation}
where
\begin{equation}
\mathbf{T}_{\pm}=\sum\limits_{k=1}^{k_{\pm}}q_k^{\pm}\,\mathbf
T^{\pm}_{2k-1,2k} ~.
\end{equation}
Thus, the connection and the source for this configuration read
\begin{eqnarray}
\mathbf{A} &=& \mathbf{A}_{\text{AdS}} + \mathbf{T}\ d\phi_{12} ~, \\ [0.8em]
\mathbf{j}_{[0]} &=& 2 \pi \alpha \left( \mathbf{J}_{12} - \mathbf{T} \right) \delta(\Sigma_{12}) ~.
\end{eqnarray}
For the Killing spinors, we already know that when $k_+ = k_- = 0$, there is no solution apart from the global AdS space, whereas for $k_+$ or $k_- \neq 0$ (say, $k_+ = 1$), the system resembles the $\mathcal{N}=2$ case, and a Killing spinor of the type (\ref{psi(2,0)}) exists for $(1-\alpha)\,(1+2q_1^{+}) \in \mathbb{Z}$.

The $\phi_{12} $-component of the Killing spinor $\epsilon^{\pm}$ equation (\ref{Killingpm}) reads
\begin{equation}
\left( \partial_{\phi_{12}} - \frac{1}{2}\,\Gamma_{0} -
\sum\limits_{k=1}^{k_{\pm}} q_k^{\pm}\,\tau_{2k-1,2k}^{\pm} \right)\
\lambda_{\pm}(\phi_{12})=0 ~,
\end{equation}
and has the general lowest supersymmetry preserving solution
\begin{equation}
\epsilon^{\pm} = \exp\left\{ \pm f(r) \pm \frac{i}{2 \ell}\,(1-\alpha)t \pm \frac{i}{2}\,(1-\alpha) \left(1 + \sum\limits_{k=1}^{k_{\pm}} q_k^{\pm} \right) \phi\right\}\, \eta^{\pm} ~.
\end{equation}
The constant spinors $\eta^{\pm}$ are chosen such that
\begin{eqnarray}
& & \left( \Gamma _{0}\right)_{\ \beta }^{\alpha}\,\eta_{I}^{\pm
\beta} = i \eta_{I}^{\pm \alpha} ~, \label{spinpro} \\ [0.8em] & &
\left( \tau_{2k-1,2k}^{\pm}\right)_{\ I}^{J}\,\eta _{J}^{\pm} = i
\eta_{I}^{\pm} ~, \quad k = 1, \ldots, k_\pm ~. \label{projk}
\end{eqnarray}
This gives raise to $p - k_+$ ($q - k_-$) independent components. The boundary condition $\mathbf{\epsilon}(\phi +2\pi )=\pm \mathbf{\epsilon}(\phi )$ leads to the condition on the charges
\begin{eqnarray}
(1-\alpha)\left( 1+q_{12}^{+}+\cdots +q_{k_{+}-1,k_{+}}^{+}\right)  &\in &\mathbb{Z} ~, \\ [0.8em]
(1-\alpha)\left( 1+q_{12}^{-}+\cdots +q_{k_{-}-1,k_{-}}^{-}\right)  &\in &\mathbb{Z} ~.
\end{eqnarray}
Notice that each projection in (\ref{projk}) effectively acts on a two-dimensional subspace because it corresponds to an Abelian rotation inside the Cartan subgroup of $osp(p|2)\times osp(q|2)$. Thus, at the beginning, there were $\mathcal{N}=p+q$ (real two-component) spinors, and $k_{+}+k_{-}$ Abelian projections leave $\mathcal{N}-(k_{-}+k_{+})$ vectorial components unchanged. Furthermore, the spinorial projection (\ref{spinpro}) breaks a half of supersymmetries, that finally gives $\left[\mathcal{N} -( k_{-} + k_{+})\right]/2$ supercharges.

So far, we have shown that the three-dimensional spacetime containing a $0$-brane admits a globally defined Killing spinor, by explicitly constructing it. This should be sufficient to guarantee this
geometry to be a stable vacuum for supersymmetric theories with different values of $\mathcal{N}$. The supersymmetry algebra establishes a lower bound for the energy, which is saturated by the vacuum configuration. Thus, it is possible to assert the stability of the purely bosonic configuration by just checking that a Killing spinor can exist in that background.

The only missing link in this proof of stability is that we have not shown the charges that satisfy the supersymmetry algebra to be defined for this configuration. In fact, the charges that generate the symmetry (super) group should be finite and satisfy the right Poisson algebra in the phase space of the theory.

In $2+1$  dimensions, it is rather straightforward to check that the canonical charges satisfy the algera of the supersymmetric extension of the AdS group, that is the super Virasoro algebra
\cite{Banados1996c,Banados1999d,Coussaert1994}. Since there is not much difference in the construction for naked singularities and for the standard black holes, we will not devote more to this discussion here. However, in the five-dimensional case that follows, the construction of the charges and the establishment of the energy lower bound will be explicitly carried out.

%%%%%%%%%%%%%%%%%%%%%%%%%%%%%%%%%%%%%%%%%%%%%%%%%%%%%%%%%%%%%%%%%%%%%%%%%%%%%%%%
\section{Static 2-brane in AdS$_{5}$  \label{D=5}}
%%%%%%%%%%%%%%%%%%%%%%%%%%%%%%%%%%%%%%%%%%%%%%%%%%%%%%%%%%%%%%%%%%%%%%%%%%%%%%%%

Now we turn to the generalization of the conical defect discussed above to higher dimensions. There are two obvious generalizations of a $0$-brane in $3$-dimensional AdS space: a \thinspace $0$-brane in AdS$_{D}$ or a $(D-3)$-brane in AdS$_{D}$.

The first one is a topological defect obtained as a surface deficit of a $(D-2)$-sphere, that produces naked singularities corresponding to the negative energy states of the black hole spectrum obtained in Chern-Simons gravities \cite{Miskovic2009}. These solutions have divergent curvature as $r\rightarrow 0$ (the space is not locally AdS), and correspond to pathological, possibly unstable geometries, which are unlikely to support Killing spinors.

The second option, where the topological defect is constructed as an angular deficit in a two-dimensional plane, gives rise to spacetimes of constant negative curvature, with point-like singularities at the fixed points of some identification Killing vector. Those singularities can be understood as the position of the $(D-3)$-brane, in close analogy with the situation discussed above.

Consider a ($D-3$)-brane in locally AdS$_{D}$ space produced by a 2-form current with support at the center of a $2$-plane (the transverse space, $\Sigma_{12}$). The ($D-3$)-brane has internal local symmetry $SO(D-3,2)$ and is invariant under external $SO(2)$ rotations in the transverse space. The full symmetry of the system is therefore $SO(2)\times SO(D-3,2)$. Adding more dimensions to the brane does not change the picture \cite{Miskovic2009}.

Focusing on the five-dimensional case, the idea is to enlarge the 0-brane in AdS$_{3}$ with two additional coordinates, $x^{3}$ and $x^{4}$,
\begin{equation}
\begin{array}{ll}
x^{0} = A\,\cos \phi _{05}\cosh \rho ~, & \qquad\qquad x^{1} = B\,\cos \phi_{12} ~, \\ [0.8em]
x^{5} = A\,\sin \phi _{05}\cosh \rho ~, & \qquad\qquad x^{2} = B\,\sin \phi_{12} ~, \\ [0.8em]
x^{3}=A\,\sinh \rho \cos \theta ~, & \qquad\qquad x^{4}=A\,\sinh \rho \sin \theta ~.
\end{array}
\label{x0-x5}
\end{equation}
Here, the functions $A\in \lbrack \ell ,\infty )$ and $B\in \lbrack 0,\infty)$ satisfy the constraint $A^{2}-B^{2}=\ell ^{2}$, ensuring that the spacetime is locally AdS$_{5}$, $\eta _{AB}\,x^{A}x^{B}=-\ell^{2}$. Introducing an angular deficit $\alpha$ in the $x^{1}$-$x^{2}$ plane,
the 2-brane sits at $B=0$ ($A=\ell $). In AdS$_{5}$, $B\in \lbrack 0,\infty)$ is the radius of the cylinder with the brane at its axis; $\phi _{12}\in [0,2\pi (1-\alpha))$ is the azimuthal angle of the cylinder; $\phi _{05}\in (-\infty ,\infty)$ is a time coordinate both in the braneworld and in the target space ($\phi _{05}$ is unwrapped in order to avoid closed timelike curves); $\rho \in \lbrack 0,\infty )$ is the internal radial coordinate in the brane; and $\theta \in \lbrack 0,2\pi )$ is an internal angular coordinate on the brane. The three-dimensional 0-brane is recovered when $\rho=0$.

Notice that the intrinsic geometry of the brane is AdS$_{3}$ and (\ref{x0-x5}) is a solution of the theory in the sector where $\mathbf{F}=\mathbf{j}_{[2]}$, where the source 2-form is
\begin{equation}
\mathbf{j}_{[2]}=2\pi \alpha \,\delta (\Sigma _{12})\,\mathbf{J}_{12} ~.
\label{gravitysource}
\end{equation}
The metric now reads
\begin{equation}
ds^{2}=\frac{\ell ^{2}dB^{2}}{B^{2}+\ell ^{2}}+B^{2}d\phi _{12}^{2} + \left( B^{2} + \ell^{2} \right) \left( d\rho ^{2}+\sinh ^{2}\rho \,d\theta ^{2}-\cosh^{2}\rho \,d\phi _{05}^{2}\right) ~.  \label{staticbrane}
\end{equation}
This can also be described in terms of a standard periodic angular coordinate $\phi \in [0,2\pi)$(without defect in the $\Sigma_{12} $\ plane) and the usual unwrapped time coordinate $t$. This is
achieved by rescaling $\left( B,\phi _{12},\phi _{05}\right) =\left( \frac{r}{1-\alpha},(1-\alpha)\phi,\frac{(1-\alpha)t}{\ell} \right)$, so the metric (\ref{staticbrane}) becomes
\begin{equation}
ds^{2}=\frac{dr^{2}}{\frac{r^{2}}{\ell ^{2}}-M^2}+r^{2}d\phi^{2} + \left(\frac{r^{2}}{\ell^{2}}-M^2\right) \left( -\cosh ^{2}\rho \,dt^{2} - \frac{\ell^{2}}{M^2}\,d\rho ^2-\frac{\ell^{2}}{M^2}\,\sinh ^{2}\rho\,d\theta ^{2}\right) ~.
\label{5D naked}
\end{equation}
Here the negative mass parameter $M = -(1-\alpha)^{2}$ characterizes a naked singularity. The three-dimensional spacetime section multiplied by the scale factor $(\frac{r^{2}}{\ell ^2}-M^2)$ and parameterized by $(t,\rho ,\theta )$, is the worldvolume of the brane with global AdS$_{3}$ geometry of radius $\ell /(1-\alpha)$.

%%%%%%%%%%%%%%%%%%%%%%%%%%%%%%%%%%%%%%%%%%%%%%%%%%%%%%%%%%%%%%%%%%%%%%%%%%%%%%%%
\subsection{BPS states \label{BPS states 5D}}
%%%%%%%%%%%%%%%%%%%%%%%%%%%%%%%%%%%%%%%%%%%%%%%%%%%%%%%%%%%%%%%%%%%%%%%%%%%%%%%%

The supersymmetric extension of the AdS group in five dimensions is $SU(2,2|N)$, whose bosonic sector is AdS$_{5}\times $ $SU(N)$ $\times $ $U(1)$. The supergroup is generated by the set $\mathbf{G}_{K}=\{\mathbf{G}_{\bar{K}},\mathbf{Z}\}$, where $\mathbf{Z}$ is the $U(1)$ generator and $\mathbf{G}_{\bar{K}}$ represent the AdS generators $\mathbf{J}_{AB}=\{\mathbf{J}_{ab}, \mathbf{P}_{a}\}$, the $SU(N)$ generators $\mathbf{T}_{\Lambda }$ ($\Lambda=1,\ldots ,N^{2}-1$) and the supersymmetry generators $\mathbf{Q}_{s}^{\alpha }$ that transforms in a vector representation of $SU(N)$ labeled by $s=1,\ldots ,N$, where $\alpha = 1, \ldots ,4$ are spinor indices. For more details, see Appendix \ref{SAdS5}.

The gauge connection 1-form that takes values in this Lie superalgebra has components
$$
\mathbf{A} = \frac{1}{\ell}\,e^{a}\,\mathbf{P}_{a} + \frac{1}{2}\,\omega ^{ab}\,\mathbf{J}_{ab} + a^{\Lambda}\,\mathbf{T}_{\Lambda} + \left( \bar{\psi}_{\alpha}^{s}\,\mathbf{Q}_{s}^{\alpha} - \mathbf{\bar{Q}}_{\alpha}^{s}\,\psi_{s}^{\alpha} \right) + b\,\mathbf{Z} ~.
$$
The associated field-strength in the bosonic sector ($\psi =0$) is
\begin{equation}
\mathbf{F} = \frac{1}{2}\,F^{ab}\,\mathbf{J}_{ab} + \frac{1}{\ell }\,T^{a}\,\mathbf{J}_{a} +\mathcal{F}^{\Lambda}\,\mathbf{T}_{\Lambda} + f\,\mathbf{Z} ~,
\label{F-components}
\end{equation}
where $F^{ab}=R^{ab}+\frac{1}{\ell ^{2}}e^{a}\wedge e^{b}$, and $T^{a}=de^{a}+\omega_{~\,b}^{a}\wedge e^{b}$ is the torsion 2-form. The $U(1)$ and $SU(N)$ field-strengths are
\begin{equation}
f=db ~,\qquad \text{and} \qquad \mathcal{F}=da+a\wedge a ~,
\end{equation}
respectively. Here $a\equiv a^{\Lambda }\tau _{\Lambda }$ is the $SU(N)$ connection and $\tau _{\Lambda }$ are the $su(N)$ generators.

The Chern-Simons field equations for the 2-brane have the form
\begin{equation}
\left\langle \mathbf{F}\wedge \left( \mathbf{F}-\mathbf{j}_{[2]}\right)
\mathbf{G}_{K}\right\rangle =0 ~,
\label{eom N}
\end{equation}
where the source $\mathbf{j}_{[2]}=j^{L}\mathbf{G}_{L}$ is a two-form with support at the center of the transverse space.

Equation (\ref{eom N}) comes from the five-dimensional version of the action (\ref{action}), which now reads
\begin{equation}
I\left[ \mathbf{A},\mathbf{j}\right] =\kappa \int_{\mathbb{M}^{5}}\langle \mathbf{C}_{5}(\mathbf{A})-\mathbf{j}_{[2]}\wedge \mathbf{C}_{3}(\mathbf{A})\rangle ~.
\label{action-5}
\end{equation}
A Killing spinor $\boldsymbol{\epsilon} = \bar{\epsilon}_{\alpha }^{s}\,\mathbf{Q}_{s}^{\alpha} - \mathbf{\bar{Q}}_{\alpha}^{s}\,\epsilon_{s}^{\alpha}$ is a solution of the equation
\begin{equation}
(D\boldsymbol\epsilon)_{s} \equiv \left[ d+\frac{1}{4}\,\omega ^{ab}\,\Gamma_{ab} + \frac{1}{2\ell}\,e^{a}\,\Gamma_{a} + i \left( \frac{1}{4}-\frac{1}{N}\right) b\right] \epsilon_{s} - a_{\;s}^{r}\,\epsilon _{r}=0 ~,
\label{Killing Spinor Eq}
\end{equation}
where $a_{\;r}^{s}=a^{\Lambda }(\tau _{\Lambda })_{\;r}^{s}$. The consistency condition for the Killing spinor equation (\ref{Killing Spinor Eq}) in the region surrounding the singular points that must be removed from the manifold, $DD\mathbf{\boldsymbol\epsilon }=[\mathbf{F},\mathbf{\boldsymbol\epsilon }]=0$, reads
\begin{equation}
\left[ \frac{1}{4}\,F^{ab}\,\Gamma_{ab} + \frac{1}{2\ell }\,T^{a}\,\Gamma_{a} + i \left( \frac{1}{4}-\frac{1}{N}\right) f\right] \epsilon_{s} - \mathcal{F}_{\;s}^{r}\,\epsilon_{r} = 0 ~,  \label{SpinorEqConsistency}
\end{equation}
where $\mathcal{F}^r_{\ s} = \mathcal{F}^\Lambda(\tau_\Lambda)^r_{\ s}$.

We are interested in the $p$-brane solutions, that is, with locally AdS spacetime, $F^{AB}=0$. Following \cite{Miskovic2006}, we choose the ansatz that restricts the $SU(N)$ field strength, $\mathcal{F}$, to the two-dimensional $r$-$\theta $ submanifold in spacetime, $\mathcal{F}^{\Lambda} = \mathcal{F}_{r\theta }^{\Lambda}\,dr\wedge d\theta$, and the magnetic part of the $U(1)$ field strength, $f_{ij}$, to be invertible. The source-free field equations become
\begin{eqnarray}
\left( \frac{1}{N^{2}}-\frac{1}{4^{2}}\right) f\wedge f &=&0 ~, \label{f_f} \\ [0.8em]
\mathcal{F}^{\Lambda }\wedge f &=&0 ~.
\label{ansatz eom}
\end{eqnarray}
As shown in \cite{Miskovic2005}, the dynamical content of the theory depends crucially on $N$. Moreover, for a given $N$, the phase space has various sectors each with a different number of degrees of freedom. In that reference it was also proven that the canonical charges can be calculated only in the \emph{canonical} sectors, where the symplectic form has maximal rank. In these sectors, the Hamiltonian analysis can be safely applied and the theory possesses maximal number of degrees of freedom. In our case, the ansatz (\ref{ansatz eom}) is canonical only if the magnetic part of the Abelian field strength, $f_{ij}$, is invertible. Invertibility requires a nonvanishing Pfaffian of $f_{ij}$, which implies $f\wedge f\neq 0$ and therefore the canonical sector we will consider only holds for $N=4$. For $D=5, N\neq4$, the invertibility of $f_{ij}$ is not required and the BPS states in these cases are treated in a way common for all $D>5$ with $\mathcal{N}>2$ supersymmetries, and will be discussed later.

For the locally AdS geometry, the consistency condition (\ref{SpinorEqConsistency}) is
\begin{equation}
\mathcal{F}_{\;s}^{r}\,\epsilon_{r}=0 ~,
\label{Consistency N=4}
\end{equation}
hence, $\mathcal{F}$ must be nonvanishing for more than one value of the index $r$, so that the contributions of all components cancel. Taking advantage of the isomorphism $su(4)\simeq so(6)$, the $SU(4)$ curvature can be expressed as $\mathcal{F}_{r}^{s}=\frac{1}{2}\mathcal{F}^{IJ}\left( \tau_{IJ}\right)_{r}^{s}$, where the $so(6)$ generators
\begin{equation}
\tau_{IJ}=\frac{1}{4}\,\left[ \hat{\Gamma}_{I},\hat{\Gamma}_{J}\right] ~, \qquad \left( I,J=1,\ldots ,6\right) ~,
\end{equation}
are given in terms of the Euclidean Dirac matrices $\hat{\Gamma}_{I}$. The commuting matrices $\tau_{12}$ and $\tau _{34}$ generate a $U(1)\times U(1)$ subgroup for $SU(4)$, and since $\left( \tau_{12}\right)^{2} = \left( \tau_{34}\right)^{2} = - \frac{1}{4}$, their eigenvalues are $\pm \frac{i}{2}$.

A simple nontrivial solution of (\ref{Killing Spinor Eq}) is a ``twisted'' configuration \cite{Edelstein1996,Edelstein1996b}, for which the only nonvanishing $U(1)\times U(1)$ components of the $SU(4)$ curvature are $\mathcal{F}^{12}=da^{12}$ and $\mathcal{F}^{34}=da^{34}$, and the Killing spinor $\mathbf{\epsilon }$ is assumed to satisfy
\begin{equation}
\left( \tau_{12}\right) _{s}^{r}\epsilon _{r}=\frac{i}{2}\,\epsilon_{s} ~,\qquad \left( \tau _{34}\right)_{s}^{r}\epsilon _{r}=-\frac{i}{2}\,\epsilon _{s} ~.
\label{twist}
\end{equation}
These {\it chirality} projections preserve $1/4$ of the supersymmetries. Therefore, the consistency condition (\ref{Consistency N=4}) becomes
\begin{equation}
\frac{i}{2}\,\left( \mathcal{F}^{12}-\mathcal{F}^{34}\right) \,\epsilon_{s}=0 ~,
\end{equation}
which is solved by $\mathcal{F}^{12}=\mathcal{F}^{34}$. Then the $SU(4)$ field has only one independent component
\begin{equation}
a^{12} = a^{34} + d\Omega (x^{i}) ~,
\label{1234}
\end{equation}
where $\Omega \left( x^{i}\right) $ is an arbitrary phase. Using the identity $a_{r}^{s}\,\epsilon_{s}=\frac{i}{2}\,d\Omega \,\epsilon_{r}$, the Killing equation (\ref{Killing Spinor Eq}) reduces to
\begin{equation}
\left( d+\frac{1}{4}\,\omega ^{ab}\Gamma_{ab} + \frac{1}{2\ell}\,\Gamma _{a} - \frac{i}{2}\,d\Omega \right)\,\epsilon_{s} = 0 ~.
\label{Killing Eq Twisted}
\end{equation}
We shall take a particular choice of $\Omega$ that produces nontrivial topology, since it introduces a new source of charge $q$,
\begin{equation}
\Omega (x) = q\,\phi_{12} ~,
\end{equation}
which represents a line of flux $2\pi q$ piercing through the center of the $\Sigma_{12}$-plane. In general, there may be other solutions with different $\Omega$'s, but this is sufficient for our purposes here. The remaining components of the $SU(4)$ and $U(1)$ gauge fields can be chosen so as to satisfy all the consistency conditions, for example
\begin{eqnarray}
\mathbf{A}_{\text{AdS}} &=&\text{Static 2-brane at the center of } \Sigma_{12} ~, \label{A_brane} \\ [0.9em]
a &=&h(\theta )\,dr\,\left( \mathbf{T}_{12}+\mathbf{T}_{34}\right) +q\,d\phi_{12}\,\mathbf{T}_{12} ~,  \label{a} \\ [0.9em]
b &=&\left[ \mathcal{B}\,g(r)\rho \,d\theta + \mathcal{E}\,r\,d\phi \,\right]
\mathbf{Z} ~.  \label{b}
\end{eqnarray}
Here $h=h(\theta)$ is an arbitrary periodic function of $\theta $, $\mathcal{E}$ and $\mathcal{B}$ are nonvanishing constants. The function $g(r)$ is continuous and satisfies the boundary conditions
\begin{equation}
g(0) = 0 ~,\qquad g(\infty) = 1 ~.
\end{equation}
These conditions are imposed to ensure that $f_{\rho\theta}$ vanishes at the singularity and is nontrivial on the boundary. One possible choice is $g(r)=r^{2}/(r^{2}+\ell ^{2})$. For $r\neq 0$ (outside the singularity), the field strength for the connection (\ref{A_brane}-\ref{b}) takes the form
\begin{equation}
\mathbf{F}_{r\neq 0} = - h^{\prime }(\theta )\,dr\wedge d\theta \,\left( \mathbf{T}_{12} + \mathbf{T}_{34} \right) + \left[ \mathcal{B\,}g^{\prime }(r)\rho \,dr\wedge d\theta + \mathcal{B\,}g(r)\,d\rho \wedge d\theta +\mathcal{E}\,dr\wedge d\phi \right]\,\mathbf{Z} ~.
\label{F}
\end{equation}
The first term in the RHS corresponds to $\mathcal{F}$, and the second to $f$. It is easily checked that $\det \left( f_{ij}\right) =\mathcal{B}^{2}\,\mathcal{E}^{2}\,g^{2}(r)\neq 0$ and $\mathcal{F}\wedge f=0$, as required by (\ref{ansatz eom}).

At $r=0$, the field strength acquires an additional term coming from the conical singularity in the 1-2 plane, $dd\phi _{12}=-2\pi \alpha \,\delta(r)\,dr\wedge d\phi $, so that the full AdS curvature is
\begin{equation}
\mathbf{F}=\mathbf{F}_{r\neq 0}+\mathbf{j}_{[2]} ~,
\label{F calculated}
\end{equation}
where a source has the form
\begin{equation}
\mathbf{j}_{[2]}=2\pi \alpha \,\left( \mathbf{J}_{12}-q\,\mathbf{T}_{12}\right) \,\delta (r)\,dr\wedge d\phi ~.
\end{equation}
There are no contributions of the conical singularity to the torsion $T_{\mu\nu}^{a}$ or to the $U(1)$ field strength $f$ because of the identity $r\delta (r)=0$. The first term in the RHS represents the topological defect related to the (negative) mass, $M = -(1-\alpha )^{2}$, and $2\pi \alpha q$ is related to the magnetic charge $q$ coming from the broken $SU(4)$ group.

However, it is not yet guaranteed that the ansatz (\ref{A_brane}-\ref{b}) that satisfies the consistency condition and the source-free equations $\left\langle \mathbf{F}_{r\neq 0}^{2}\,\mathbf{G}_{K}\right\rangle = 0$, is also a solution of the full field equations (\ref{eom N}). To check this, one may write Eq.(\ref{F calculated}) in the equivalent form
\begin{equation}
\left\langle \left( \mathbf{F}-\mathbf{F}_{r\neq 0}-\mathbf{j}_{[2]}\right)^{2}\mathbf{G}_{K}\right\rangle = 0 ~.
\end{equation}
Then, using the fact that $\mathbf{j}_{[2]}\wedge \mathbf{j}_{[2]}\equiv 0$ (because $\mathbf{j}_{[2]}$ is defined on a two-dimensional plane), and since $\mathbf{F}_{r\neq 0} \wedge
\mathbf{j}_{[2]}$ is proportional to $g(0)=0$, one can see that equation (\ref{eom N}) is indeed satisfied.

Knowing the background configuration, we can solve equation (\ref{Killing Eq Twisted}), that has the same form as the Killing equation in pure Chern-Simons AdS gravity (see Appendix \ref{pure 5D 2-brane}), up to the $U(1)$ gauge function $\Omega =q\phi _{12}$ that gives an additional shift to the exponent,
\begin{equation}
\epsilon _{s}=e^{-f(r)\Gamma _{1}}e^{-\frac{1}{2}\,\rho \,\Gamma _{3}}e^{\frac{1}{2}\,\phi _{12}\,\Gamma _{12}}e^{\frac{i}{2}\,q\phi _{12}}e^{-\frac{1}{2}\,\phi _{05}\Gamma _{0}}e^{\frac{1}{2}\,\theta \Gamma _{34}}\eta_{s} ~.
\label{Killing spinor 2brane}
\end{equation}
The constant spinors $\eta _{s}$ are chosen as common eigenvectors of the commuting $su(4)$ generators,
\begin{equation}
\Gamma_{12}\,\eta _{s}=-i\eta_{s} ~, \qquad \Gamma_{34}\,\eta_{s}=-i\eta_{s} ~.
\label{projections}
\end{equation}
Because $\Gamma _{0}$ also commutes with $\Gamma _{12}$ and $\Gamma _{34}$ (it is normalized as $\Gamma_{0}=i\Gamma_{12}\Gamma _{34}$, see Appendix \ref{SAdS5}), we additionally have
\begin{equation}
\Gamma_{0}\,\eta _{s}=-i\eta _{s} ~.
\end{equation}
Taking these projections into account, we can analyze global properties of the fermion $\epsilon _{s}$. Requiring that it is single-valued under $\phi \rightarrow \phi +2\pi $ and $\theta \rightarrow \theta +2\pi $ that is, with periodic (Ramond (R) sector) or anti-periodic (Neveu-Schwartz (NS) sector) boundary conditions, the charge $q$ becomes quantized,
\begin{equation}
(1-\alpha)\left( q-1\right) =n\in \mathbb{Z} ~.
\label{quantization}
\end{equation}
We finally obtain the Killing spinor in the form
\begin{equation}
\epsilon_{s}=e^{-f(r)\Gamma_{1}}e^{-\frac{1}{2}\,\rho \Gamma _{3}}e^{\frac{i}{2}\,\left( n\phi -\theta \right) }e^{\frac{i}{2\ell}\,(1-\alpha)t}\,\eta_{s} ~.
\end{equation}
To calculate the number of preserved supersymmetries, we recall that the group projective operators (\ref{twist}) already break $3/4$ of the supercharges. These are complemented by the spinorial {\it chirality} projections (\ref{projections}) that analogously preserve $1/4$ of the remaining supersymmetries. All in all, this finally gives a $1/16$ BPS state. There are two unbroken
supercharges, which correspond to the minimum number compatible with the expected symmetries of a $2$-brane worldvolume. In the next section, we will show that for $N \neq 4$, there exists a fivedimensional 2-brane that is a $1/4$ BPS state.

The remaining bosonic symmetry of this BPS solution is described by the Killing vectors $\lambda ^{K}$ ($D\lambda ^{K}=0$), and it includes the original $U(1)$ symmetry with the parameter $\lambda ^{Z}$, as well as the Cartan subgroup $U(1)\times U(1)\times U(1)$ of $SU(4)$ (where the third  $U(1)$ generator is not switched on in our configuration, though) with the gauge parameters $\lambda ^{12}$, $\lambda^{34}$ and $\lambda^{56}$. The isometries of the 2-brane are linear combinations of $\partial_{t}$ and $\partial_{\phi}$. The proof is given in Appendix \ref{CovConstant}.

In the considered BPS solution, we removed the singularity from the manifold when solving the consistency of the Killing Spinor equation (\ref{SpinorEqConsistency}). If, however, one were to insist on including the singular point, one would have to take into account the contribution of the
sources. Using the twisting of the $SU(4)$ field (\ref{twist}), as well as the AdS projections (\ref{projections}), one obtains
\begin{equation}
2\pi \alpha\,\frac{i}{2}\left( 1+\frac{q}{2}\right) \,\delta (\Sigma_{12})\,\epsilon _{s} = 0 ~.
\end{equation}
Clearly, when we have the singularity cut from the manifold this equation does not arise. Otherwise, we have to integrate this expression over the transverse section $\Sigma_{12}$, and we obtain
\begin{equation}
q = -2 ~,
\end{equation}
meaning that the electromagnetic charge has a fixed value. From (\ref{quantization}), this also implies that
\begin{equation}
\alpha=1-\frac{|n|}{3} ~,\qquad |n|=1,\,2,\,3 ~.
\label{n/3}
\end{equation}
Note that $n=\pm 3$ corresponds to global AdS$_{5}$, while $|n|=1,2$ are two different charged BPS 2-branes.

%%%%%%%%%%%%%%%%%%%%%%%%%%%%%%%%%%%%%%%%%%%%%%%%%%%%%%%%%%%%%%%%%%%%%%%%%%%%%%%%
\subsection{Canonical charges and their central extensions}
%%%%%%%%%%%%%%%%%%%%%%%%%%%%%%%%%%%%%%%%%%%%%%%%%%%%%%%%%%%%%%%%%%%%%%%%%%%%%%%%

In canonical sectors of Chern-Simons theories, the splitting between first and second class constraints can be performed explicitly \cite{Banados1996a} and the conserved charges can be calculated following the Regge-Teitelboim approach \cite{Regge1974},
\begin{equation}
Q\left[ \lambda \right] =-\kappa \int_{\Sigma_{\infty }}\left\langle \mathbf{\lambda\, \bar{F}}\wedge \mathbf{A}\right\rangle ~,
\label{charges}
\end{equation}
where $\Sigma_{\infty }$ is the boundary at spatial infinity ($r\rightarrow \infty $), $\mathbf{\lambda} = \lambda^{K}(x)\,\mathbf{G}_{K}$ is a gauge parameter and $\mathbf{\bar{F}}$ is the background field strength. The charge is obtained assuming the boundary conditions $\mathbf{A}\rightarrow \mathbf{\bar{A}}$ and $\mathbf{\lambda} \rightarrow \mathbf{\bar{\lambda}}$, where $\mathbf{\bar{\lambda}}$ are asymptotic Killing vectors of the background, $\bar{D}\mathbf{\bar{\lambda}}=0$. Let us emphasize that the bar denotes the \textit{spatial
asymptotic} sector of the solutions.

The algebra of charges generically picks up a central extension of the gauge algebra \cite{Brown1986a},
\begin{equation}
\left\{ Q\left[ \lambda \right] ,Q\left[ \eta \right] \right\} =Q\left[ \left[ \lambda ,\eta \right] \right] +C\left[ \lambda ,\eta \right] ~,
\label{charge algebra}
\end{equation}
where $\left[ \lambda ,\eta \right] ^{K}=f_{MN}^{\ \ \ \ K}\,\lambda^{M}\eta ^{N}$, and the central charge $C\left[ \lambda ,\eta \right] $ has the form
\begin{equation}
C\left[ \lambda ,\eta \right] =-\kappa \int_{\Sigma_{\infty}} \left\langle \mathbf{\lambda\, \bar{F}}\wedge d\mathbf{\eta }\right\rangle ~.
\label{general C}
\end{equation}

In a locally AdS spacetime with a 2-brane whose metric is given by (\ref{staticbrane}), the spatial boundary $\Sigma _{\infty }$ taken at constant radius $B=r/(1-\alpha)\rightarrow \infty $, and constant time, $\phi _{05}=(1-\alpha)\,t/\ell = const$, the asymptotic metric reads
\begin{equation}
\left. ds^{2}\right\vert_{\Sigma_{\infty }}=\frac{r^{2}}{(1-\alpha)^{2}}\,\left( d\rho ^{2} + \sinh^{2}\rho\, d\theta^{2} \right) + r^{2}d\phi^{2} ~,
\end{equation}
and its topology is isomorphic to $H^{2}\times S^{1}$, where $H^{2}$ is a two-dimensional hyperboloid parameterized by $(\rho ,\theta )$.

Evaluating eq.(\ref{F}) for our case of $SU(4)\times U(1)$ matter at the boundary ($\mathbf{j}=0$ and $dr=0$), gives the Abelian background field strength on $H^{2}$,
\begin{equation}
\mathbf{\bar{F}}=\mathcal{B}\,d\rho \wedge d\theta \,\mathbf{Z} ~.
\label{F background}
\end{equation}
Around this background, the charges (\ref{charges}) are
\begin{equation}
Q\left[ \lambda \right] =\int_{\Sigma_{\infty }}d^{3}x\,\lambda^{K}q_{K} ~, \qquad q_{K} =\frac{\kappa\, \mathcal{B}}{4}\,\gamma_{KL}\,A_{\phi}^{L} ~,
\label{Q evaluated}
\end{equation}
where $\gamma_{KL}\,$is the invariant Killing bilinear form for $PSU(1,1|4)$ (see Appendix \ref{SAdS5}), and the normalization of the volume element $d^{3}x$ of $\Sigma _{\infty }$ has been chosen as $\frac{d\phi }{2\pi}$ for $S^{1}$ and $\frac{d\rho \,d\theta }{\text{Vol}(H^{2})}$ for $H^{2}$. Note that the class of solutions considered here has identically vanishing $U(1)$ charge (generated by $\mathbf{Z}$), since $\gamma _{zz}=\gamma _{\bar{K}z}=0$.

At infinity, the gauge parameter $\mathbf{\lambda }$ approaches $\mathbf{\bar{\lambda}}$, a covariantly constant vector that describes asymptotic symmetries of the background configuration. It possesses a $U(1)$ symmetry associated to $\bar{\lambda}^{Z}=const$, $U(1)\times SO(4)$ symmetry coming from $SU(4)$ gauge parameters $\bar{\lambda}^{12}=const$ and $\left. \bar{\lambda}^{IJ}\right\vert _{I,J\in \{3,4,5,6\}}=Const$, and the only nonvanishing AdS parameters are $\bar{\lambda}^{25} = -\bar{\lambda}^{12} = const.$ describing a $\partial_{\phi }$ isometry of the 2-brane. Derivation of $\bar{\lambda}^{K}$ is given in Appendix \ref{CovConstant}.

Splitting the connection into the background AdS and the rest,
\begin{equation}
\mathbf{\bar{A}} = \mathbf{\bar{A}}_{\text{AdS}}+\left( (1-\alpha)\,q\,\mathbf{T}_{12} + \mathcal{E}\, r\,\mathbf{Z}\right) \,d\phi +\mathcal{B}\, \rho \,\mathbf{Z}\,d\theta ~, \qquad (1-\alpha)(q-1)=n\,\ (n\in \mathbb{Z}) ~,
\label{A background}
\end{equation}
where the AdS part includes a static 2-brane (see Appendix \ref{pure 5D 2-brane}),
\begin{eqnarray}
&& \mathbf{\bar{A}}_{\text{AdS}} = \frac{r}{\ell }\,\left( \mathbf{J}_{25} - \mathbf{J}_{12}\right) \,d\phi +\frac{r}{(1-\alpha)\ell }\,\left( \mathbf{J}_{35} - \mathbf{J}_{13}\right) \,d\rho \nonumber \\ [1em]
&& \qquad \qquad \qquad + \left( \frac{r}{(1-\alpha)\ell }\,\sinh \rho \,\left( \mathbf{J}_{45}  -\mathbf{J}_{14}\right) -\cosh \rho \,\mathbf{J}_{34}\right) \,d\theta ~.
\label{AdS infinity}
\end{eqnarray}
Thus, the charge for the background $\mathbf{\bar{A}}$ becomes
\begin{equation}
\bar{Q}=-\kappa \int_{\Sigma _{\infty }}\left\langle \mathbf{\bar{\lambda}\,\bar{F}}\wedge \mathbf{\bar{A}}\right\rangle =\frac{\kappa\, \mathcal{B}}{4}\,(1-\alpha)\,q\,\bar{\lambda}_{su(4)}^{12} ~,
\label{Q_bgrd}
\end{equation}
which shows that this state is magnetically charged, except for the trivial case $(1-\alpha)\,q = n+1-\alpha=0$, that corresponds to global AdS$_{5}$\ ($\alpha=0$ and $n=-1$).

Furthermore, the central extension (\ref{general C}) of the algebra $psu(2,2|4)$ for this background takes the form
\begin{equation}
C\left[ \lambda ,\eta \right] = \frac{\kappa\, \mathcal{B}}{4} \int_{\Sigma _{\infty }}d^{3}x\,\gamma _{KL}\,\lambda ^{K}\partial_{\phi }\eta ^{L} ~.
\label{C explicitly}
\end{equation}
In particular, $C\left[ \lambda ^{Z},\eta ^{K}\right] \equiv 0$,\ which implies that there is no $u(1)$ central extension.

%%%%%%%%%%%%%%%%%%%%%%%%%%%%%%%%%%%%%%%%%%%%%%%%%%%%%%%%%%%%%%%%%%%%%%%%%%%%%%%%
\subsection{BPS bound} \label{BPS bound summary}
%%%%%%%%%%%%%%%%%%%%%%%%%%%%%%%%%%%%%%%%%%%%%%%%%%%%%%%%%%%%%%%%%%%%%%%%%%%%%%%%

In order to establish the existence of a BPS bound, the standard procedure is to evaluate the
anticommutator of the supercharges in the state that admits globally defined Killing spinors. The anticommutator of supersymmetry generators being positive semi-definite, leads to an inequality among the generator for bosonic symmetries. Since one of those generators is the Hamiltonian, the inequality can be used to establish a lower bound for the energy, the BPS bound, as shown in detail in Appendix \ref{BPS bound}.

The first step in this program is to write the charge (\ref{Q_bgrd}) in Fourier modes of the spatial section of the vectors $X=\left\{ \bar{\lambda}^{K},q_{K}\right\}$
\begin{equation}
X_{wmk}(r)=\int \frac{d\rho\, d\theta\, d\phi }{L\left( 2\pi \right) ^{2}}\,X(r,\rho ,\theta ,\phi )\,e^{-\frac{2\pi i}{L}\,w\rho -im\theta -ik\phi} ~.
\end{equation}
The boundary fields are functions of the periodic coordinate $\phi \in \lbrack 0,2\pi ]$ and two hyperbolic coordinates, $\rho \in \mathbb{R}$ and $\theta \in \lbrack 0,2\pi ]$. The expansion on $S^{1}\times H^{2}$ has a discrete series for $\phi$, $\theta$ and a continuous Fourier spectrum for
the non-compact coordinate $\rho $. Here we have taken $\rho \in \left[ -\frac{L}{2},\frac{L}{2}\right] $, so that Vol$(H^{2})=2\pi L$ in the limit $L\rightarrow \infty $. The bosonic modes are periodic in $\phi $, $\theta $ and therefore the numbers $m$, $k$ must be integers. For the fermionic modes these numbers can be integers and half-integers for Ramond (R) and Neveu-Schwartz (NS) boundary conditions, respectively. This gives rise to four possible sectors R$_{1}$R$_{2}$,
R$_{1}$NS$_{2}$, etc. Since $\rho $ is a noncompact coordinate, the spectrum of $w$ is continuous.

The notation is simplified calling $\vec{s}=(w,m,k)$, and consequently $\sum\limits_{\vec{s}}=\int dw\sum\limits_{m,k}$, $\delta _{\vec{s},\vec{s}^{\prime }}=\delta (w-w^{\prime })\delta _{mm^{\prime }}\delta _{kk^{\prime}}$, etc. Then, the mode expansions for the canonical and central charges (\ref{Q evaluated}) and (\ref{C explicitly}) read
\begin{eqnarray}
&& Q\left[ \lambda \right] = \sum_{\vec{s}}\lambda_{\vec{s}}^{K}\,q_{K,-\vec{s}} ~,\qquad q_{K,\vec{s}}=\frac{\kappa \mathcal{B}}{4}\,\gamma _{KL}\,A_{\phi ,\vec{s}}^{L} ~, \\ [1em]
&& C\left[ \lambda ,\eta \right] = \frac{i\kappa \mathcal{B}}{4}\,\gamma_{KL}\sum_{\vec{s},\vec{s}^{\prime }}\lambda _{\vec{s}}^{K}\,\eta_{\vec{s}^{\prime}}^{L}\,k\,\delta_{\vec{s}+\vec{s}^{\prime},0} ~,
\end{eqnarray}
and the algebra adopts the form
\begin{equation}
\left\{ q_{K,\vec{s}},q_{L,\vec{s}^{\prime }}\right\} = f_{KL}^{\quad \;M}q_{M,\vec{s} + \vec{s}^{\prime}} + \frac{i\kappa \mathcal{B}}{4}\,k\,\gamma_{KL}\,\delta _{\vec{s} + \vec{s}^{\prime},0} ~.
\label{mode algebra}
\end{equation}
This is a supersymmetric extension of the WZW$_{4}$ algebra with a nontrivial central extension for
$psu(2,2\left\vert 4\right. )$ which depends only on the $u(1)$ flux determined by $\mathcal{B}$. The modes $q_{K,\vec{s}}$ with $\vec{s}=(0,0,k)$ form a Kac-Moody subalgebra with the central charge $\kappa \mathcal{B}/4$, while the modes with $\vec{s}=(w,0,0)$ and $(0,m,0)$ form Kac-Moody subalgebras without central charges.

For the supersymmetric charges, using $q_{Z} \equiv 0$ and $\bar{q} = q^{\dagger}\,\Gamma_{0}$, the algebra (\ref{mode algebra}) multiplied by $\Gamma^{0}$ gives a positive semidefinite operator,
$$
\left\{ q_{r,\vec{s}}^{\alpha },q_{\beta ,\vec{s}^{\prime }}^{\dagger s}\right\}  = -\frac{1}{2}\,\delta_{r}^{s}\left( \Gamma ^{a}\Gamma ^{0}\right) _{\beta }^{\alpha }\,q_{a,\vec{0}}+\frac{1}{4}\,\delta_{r}^{s}\left( \Gamma ^{ab}\Gamma ^{0}\right) _{\beta }^{\alpha }\,q_{ab,\vec{0}} - \frac{1}{4}\,\left( \Gamma ^{0}\right) _{\beta }^{\alpha }\,\left( \hat{\Gamma}^{IJ}\right) _{r}^{s}\,q_{IJ,\vec{0}}-\frac{i\kappa \mathcal{B}}{4}\,k\,\left( \Gamma ^{0}\right) _{\beta }^{\alpha }\,\delta_{r}^{s} ~,
$$
which means that its eigenvalues are all positive or zero. Identifying the energy $E=q_{0,\vec{0}}$ with the time component of the AdS boost charge $q_{a,\vec{0}}=(E,q_{\bar{a},\vec{0}})$, leads to the bound (see Appendix \ref{BPS bound}),
\begin{equation}
E\geq \max \left\{ v_{\pm }+\eta _{\pm }+\frac{\kappa }{4}\,\left\vert \mathcal{B}k\right\vert \right\} ~.
\label{E}
\end{equation}
The BPS configuration has only the $SU(4)$ charge $q_{12,\vec{0}}=\frac{\kappa \mathcal{B}}{4}\,\left( 1-\alpha +n\right) \,$(see Eq.(\ref{Q_bgrd})) and the corresponding BPS energy saturates the bound,
\begin{equation}
E_{\text{BPS}}=\frac{\kappa \left\vert \mathcal{B}\right\vert }{4}\,\left( \left\vert 1-\alpha +n\right\vert + \left\vert k\right\vert \right) ~.
\end{equation}
The minimal energy corresponds to the R$_{2}$ sector $k_{\min }=0$ (no winding around the singularity)\ and in the $U(1)$ sector, we can have $n_{\min }=0$. Thus, minimal energy is
\begin{equation}
E_{\text{BPS}}^{\min }=\frac{\kappa \left\vert \mathcal{B}\,(1-\alpha )\right\vert }{4} > 0 ~.
\end{equation}
This stable configuration with nontrivial winding numbers is not the lowest energy state. As shown below, another BPS state exists for which all charges are zero and $E_{\text{BPS}}^{\min}=0$.

%%%%%%%%%%%%%%%%%%%%%%%%%%%%%%%%%%%%%%%%%%%%%%%%%%%%%%%%%%%%%%%%%%%%%%%%%%%%%%%%
\subsection{Different BPS solution}
%%%%%%%%%%%%%%%%%%%%%%%%%%%%%%%%%%%%%%%%%%%%%%%%%%%%%%%%%%%%%%%%%%%%%%%%%%%%%%%%

Consider now an ansatz where $\mathcal{F}^{\Lambda }=\mathcal{F}_{r\phi}^{\Lambda }\,dr\wedge d\phi $ and $f\wedge dr\wedge d\phi =0$, and $f_{ij}$ is invertible. Proceeding in the same way as before, the gauge connections are found to be of the form
\begin{eqnarray}
a &=& h(\phi )dr\,\left( \mathbf{T}_{12}+\mathbf{T}_{34}\right) +q\,d\phi_{12}\,\mathbf{T}_{12} ~,  \label{a2} \\ [0.9em]
b &=& \mathcal{E}\,r\,d\theta + \mathcal{B}\,\rho\,g(r)\,d\phi , \label{b2}
\end{eqnarray}
where the continuous function $g(r)$\ satisfies the boundary conditions $g(0)=0$, $g(\infty) = 1$, and the field strength for $r\neq 0$ becomes
\begin{equation}
\mathbf{F}_{r\neq 0} = h^{\prime }(\phi )\,d\phi \wedge dr\,\left( \mathbf{T}_{12} + \mathbf{T}_{34}\right) + \left( \mathcal{E}\,dr\wedge d\theta +\mathcal{B}g(r)\,d\rho \wedge d\phi
+ \mathcal{B}\rho g^{\prime }(r)\,dr\wedge d\phi \right) \,\mathbf{Z} ~.
\label{F evaluated}
\end{equation}
In (\ref{a2}) we have again included a two-brane in the implicit assumption that the angle $\phi$ has a deficit $2\pi \alpha $. Therefore the source sitting at $r=0$ is
\begin{equation}
\mathbf{j}_{[2]}=2\pi \alpha \,\left( \mathbf{J}_{12}-q\,\mathbf{T}_{12}\right) \,\delta (\Sigma _{12}) ~.
\end{equation}
On the other hand, the Abelian field $b_{\mu }$ does not contribute to the source, again because $g(0)=0$.

Since this solution is also in the canonical sector of the phase space as the previous one, and has the same sources, the Killing spinor and the relations (\ref{Killing spinor 2brane}-\ref{n/3}) remain the same. What has changed is the explicit form of the fields, Eqs. (\ref{a2}-\ref{F evaluated}), and therefore the values of the charges and central extension.

Indeed, for this solution, the background connection and the field-strength read
\begin{eqnarray}
\mathbf{\bar{A}} &=& \mathbf{\bar{A}}_{\text{AdS}}+\left( -aq\,\mathbf{T}_{12}+\mathcal{B}\rho \,\mathbf{Z}\right) \,d\phi ~, \label{alternative-A} \\ [0.9em]
\mathbf{\bar{F}} &=& \mathcal{B}\,d\rho \wedge d\phi \,\mathbf{Z} ~,
\label{alternative-F}
\end{eqnarray}
and the charges and central charge have the form
\begin{eqnarray}
Q\left[ \lambda \right] &=&\frac{\kappa \mathcal{B}}{4}\int_{\Sigma_{\infty }}d^{3}x\,\lambda ^{K}A_{\theta }^{L}\,\gamma _{KL} ~, \\ [1em]
C\left[ \lambda ,\eta \right] &=&\frac{\kappa \mathcal{B}}{4} \int_{\Sigma _{\infty }}d^{3}x\,\lambda ^{K}\partial _{\theta }\eta^{L}\gamma _{KL} ~.
\end{eqnarray}
Plugging the solution in the formula, all the charges for this configuration vanish,
\begin{equation}
\bar{Q} = 0 ~.
\end{equation}

The algebra of charges in this case has the form
\begin{equation}
\left\{ q_{K,\vec{s}},q_{L,\vec{s}^{\prime }}\right\} =f_{KL}^{\quad\;M}q_{M,\vec{s} + \vec{s}^{\prime }}+\frac{i\kappa \mathcal{B}}{4}\,m\,\gamma_{KL}\,\delta _{\vec{s} + \vec{s}^{\prime },0} ~,
\end{equation}
and a nontrivial central extension for $psu(2,2|4)$ appears only in the modes with $\vec{s}=(0,m,0)$ that correspond to the angle $\theta $. It can be shown, similarly as in the previous case, that the energy is always nonegative
\begin{equation}
E \geq \sum_{a<b}\left\vert q_{ab,\vec{0}}\right\vert +\sum_{I<J}\left\vert q_{IJ,\vec{0}}\right\vert +\left\vert \frac{\kappa \mathcal{B}}{4}\,m\right\vert ~,
\end{equation}
and the bound is saturated for the BPS state
\begin{equation}
E_{\text{BPS}} = 0 ~,
\end{equation}
where all charges, including the energy, vanish. Note that here the background is uncharged and with zero energy --that is the ground state.

This configuration is therefore a vacuum state, which is, however, not equivalent to the trivial vacuum $\mathbf{A}=0$. The main difference between thes two configurations being: while $\mathbf{A}=0$ is maximally (super) symmetric under the entire $psu(2,2|4)$ gauge algebra, whereas the ansatz (\ref{alternative-A}) is only invariant under a residual symmetry whose bosnic sector is $u(1)\times so(2,2)\times so(2)$. Moreover, the ansatz (\ref{alternative-A}) is a generic background which has propagating degrees of freedom, while the background $\mathbf{A}=0$ allow no propagation of local
perturbative excitations (it is a maximally degenerate background).

%%%%%%%%%%%%%%%%%%%%%%%%%%%%%%%%%%%%%%%%%%%%%%%%%%%%%%%%%%%%%%%%%%%%%%%%%%%%%%%%
\section{Static codimension-two branes in AdS$_{D}$}
%%%%%%%%%%%%%%%%%%%%%%%%%%%%%%%%%%%%%%%%%%%%%%%%%%%%%%%%%%%%%%%%%%%%%%%%%%%%%%%%

In the previous sections we discussed in details 0- and 2-branes in three and five dimensions, respectively, showing that these branes are stable when charged CS matter is added. These examples already exhibit the most characteristic features of higher-dimensional codimension-two branes in AdS$_{D}$.

In the introduction of Section \ref{D=5}, possible generalizations of $p$-branes to higher dimensions were discussed. We are interested here in $p$-branes living in a spacetime with constant negative curvature everywhere, except in the points where the $p$-brane is placed. These locally AdS$_D$ spacetimes are obtained by identifications of points in the global AdS$_D$, that do not change the local metric structure, but introduce topological defects. In particular, a codimension-two brane ($p=D-3$) is produced by identification of the points in a 2-plane using the Killing vector describing azimuthal rotations, $\partial _\phi$, as already discussed in three and five dimensions.

In order to construct a $(D-3)$-brane that satisfies the CS equations of motion
\begin{equation}
\left\langle \mathbf{F}^{\frac{D-3}{2}}\wedge \left( \mathbf{F}-\mathbf{j}_{[D-3]}\right) \mathbf{G}_{K}\right\rangle = 0 ~,
\label{eomD}
\end{equation}
we consider the flat embedding space $\mathbb{R}^{D-1,2}$ with the signature $(-,+,\cdots ,+,-)$ and global coordinates $x^{A}$ ($A=0,\ldots ,D$). A $(D-3)$-brane is then obtained by introducing an angular deficit $\alpha$ in the 1-2 plane $\Sigma _{12}$ parameterized by $B=r/(1-\alpha)\in \lbrack 0,\infty)$ and $\phi_{12}=(1-\alpha)\phi \in \lbrack 0,2\pi (1-\alpha))$. This identification is
generated by the Killing vector
\begin{equation}
\xi =2\pi \alpha \,(x^{2}\partial_{1}-x^{1}\partial_{2})=-2\pi \alpha\,\partial _{\phi } ~,
\end{equation}
and a conical singularity is formed at $r=0$. The coordinates $x^{0},x^{D}$ label another Euclidean plane with the radius $A=\sqrt{B^{2}+\ell ^{2}}\geq \ell$ and the angle $\phi _{0D}=(1-\alpha)\,t/\ell \in (-\infty ,\infty )$ that, unwrapped, represents time. The rest of the coordinates are introduced to have $x^{A}$ satisfying the AdS$_{D}$ constraint, $x\cdot x=-\ell ^{2}$. An explicit coordinate transformation is given in Appendix \ref{killingDdim}. Apart from the coordinates $t,r,\phi $, the metric depends on $(D-3)/2$ noncompact coordinates $\rho_{u}\in \lbrack 0,\infty)$ ($u=1,\ldots ,\frac{D-3}{2}$) that are radii of some cylinders in different directions of
$\mathbb{R}^{D-1,2}$, and also $(D-3)/2$ azimuthal angles $\phi_{2u+1,2u+2}\in \lbrack 0,2\pi )$ of these cylinders, associated to the planes $x^{2u+1}$-$x^{2u+2}$. The metric has the form that resembles the lower-dimensional cases (\ref{ds2}, \ref{5D naked}),
\begin{equation}
ds^{2}=\frac{dr^{2}}{\frac{r^{2}}{\ell ^{2}}+(1-\alpha)^{2}}+r^{2}d\phi ^{2}+\left(
\frac{r^{2}}{\ell ^{2}}+(1-\alpha)^{2}\right) d\Omega _{\text{AdS}_{D-2}}^{2} ~,
\label{D-3brane}
\end{equation}
describing a naked singularity with $M = -(1-\alpha)^{2}$, and $d\Omega _{\text{AdS}_{D-2}}^{2}$ is the global AdS$_{D-2}$ with radius $\ell /(1-\alpha)$.

The spacetime of a $(D-3)$-brane has constant negative curvature, i.e., vanishing AdS curvature $\mathbf{F}_{r\neq 0}=0$, with a point-like singularity given by the external current
\begin{equation}
\mathbf{j}_{[D-3]} = 2\pi \alpha \,\delta (\Sigma_{12})\,\mathbf{J}_{12} ~.
\end{equation}

Similarly to the lower-dimensional cases, we can ask whether this brane is stable in the framework of the CS supergravity with the superalgebra $\mathcal{G}$. The smallest superalgebras that contain the AdS$_{D}$ in the bosonic sector in odd dimensions $D\geq 5$ are given in Ref.\cite{Troncoso1999}, where it was shown that for $D=8k-1$, $8k+3$ and $4k+1$, the corresponding superalgebras $\mathcal{G}$ are $osp(N|m)$, $osp(m|N)$ and $su(m|N)$, respectively, where $m=2^{[D/2]}$. In addition to the AdS$_{D}$ generators, $\mathcal{G}$ contains the internal subalgebras $so(N)$, $sp(N)$ or $su(N)$ spanned by $\mathbf{T}^{\Lambda }$, plus some bosonic generators $\mathbf{Z}^{\Delta}$ required by the closure of the superalgebra. The spinor $\psi_{s}$ has $m$ spinorial components and it transforms as a vector under action of \ the internal group, so that there are always $\mathcal{N}=N$ gravitini, independently on the number of the bosonic generators $\mathbf{Z}^{\Delta}$.

The superalgebra-valued connection 1-form and the corresponding field-strength are given by Eqs. (\ref{generalA}) and (\ref{generalF}). In Appendix \ref{killingDdim} it is proven that the $(D-3)$-branes in the absence of CS ``matter'' ($\mathcal{A}=0$, $\psi _{s}=0$) are unstable, by showing that the Killing spinor $\epsilon _{s}$ that counts unbroken supersymmetries of the brane has the form
\begin{equation}
\epsilon _{s}=e^{-f(r)\Gamma _{1}}\prod_{u=1}^{(D-3)/2}e^{-\frac{1}{2}\,\rho_{u}\Gamma _{2u+1}}\prod_{v=1}^{(D-3)/2}e^{\frac{i}{2}\,\phi_{2v+1,2v+2}}\,e^{\frac{i}{2}\,(1-\alpha)\left( \phi -\frac{t}{\ell }\right) }\eta_{s} ~,
\label{KillingAdS}
\end{equation}
where $\Gamma_{a}$ are the $D$-dimensional $\Gamma$-matrices. $\eta_{s}$ are constant spinors that are common eigenvectors of the commuting set of matrices
\begin{equation}
\Gamma _{2u+1,2u+2}\,\eta _{s}=i\eta _{s} ~,\qquad \left( u=1,\ldots ,\tfrac{D-3}{2}\right) ~.  \label{D-projections}
\end{equation}
These projections reduce the number of supercharges by a factor $1/2^{\frac{D-3}{2}}$. In order to have the Killing spinor $\epsilon _{s}$ globally well-defined, it has to be periodic or antiperiodic under the change of all angles for a period $2\pi $, that leads to the conditon
\begin{equation}
\alpha=0 ~,
\label{a=1}
\end{equation}
which gives global AdS$_{D}$ as the only solution admitting unbroken symmetries.

In order to have a BPS codimension-two brane, one has to add other gauge fields and it is expected, based on the lower-dimensional experience, that this would permit the existence of the BPS branes in higher dimensions as well. To this end, we restrict to a class of codimension-two branes treated generically. This means that all special cases are excluded in this section. For example, we assume that $N\neq m$, because for $N=m$, the equations of motion and the dynamic content of CS gravities
take a singular form \cite{Troncoso1999}, and must be treated separately. Discussion that follows,
therefore, also covers 2-branes in $D=5$ with $N\neq 4$ matter, that were omitted in Section \ref{BPS states 5D}.

Under these assumptions, we present the simplest matter that admits nontrivial Killing spinors on a static $(D-3)$-brane: Abelian pure gauge field associated to some generator $\mathbf{T}_{1}$. The arguments that prove it go similarly as in five-dimensional case, so we will skip the details. The gauge connection is
\begin{eqnarray}
\mathbf{A}_{\text{AdS}} &=&\text{Static\thinspace (}D-3\text{)-brane\thinspace at\thinspace the\thinspace center\thinspace of\thinspace} \Sigma _{12} ~, \\ [0.9em]
\mathcal{A} &=&q\,d\phi _{12}\,\mathbf{T}_{1} ~,  \label{point-like}
\end{eqnarray}
where $\mathbf{A}_{\text{AdS}}$ is given by Eqs.(\ref{e}-\ref{w}), and the constant $q$ is magnetic charge. The field-strenght is then
\begin{equation}
\mathbf{F}=\mathbf{j}_{[D-3]} ~,
\end{equation}
where the Abelian generator contributes to the source,
\begin{equation}
\mathbf{j}_{[D-3]}=2\pi \alpha \left( \mathbf{J}_{12}-q\mathbf{T}_{1}\right) ~.
\end{equation}
This configuration clearly satisfies the equations of motion (\ref{eomD}).

For the Killing spinor $\epsilon _{s}$, the consistency condition $\left( \mathbf{F}\right) _{s}^{r}\epsilon_{r}=0$ is identically satisfied for $r\neq 0$. Nontrivial solution for $\epsilon _{s}$ is obtained similarly as in the pure gravity case, but now we also need the constant spinor to be
an eigen-vector of the Abelian generator $\mathbf{T}_{1}$,
\begin{equation}
\left( \mathbf{T}_{1}\right) _{s}^{\ r}\,\eta _{r}=-i\eta _{s} ~,
\label{T-eta}
\end{equation}
which further breaks $1/2$ of the remaining supercharges. Then, the equation $D_{\phi}\epsilon_{s}=0$ has an additional constant shift $iaq$, leading to the result
\begin{equation}
\epsilon _{s}=e^{-f(r)\Gamma _{1}}\prod_{u=1}^{(D-3)/2}e^{-\frac{1}{2}\,\rho_{u}\Gamma _{2u+1}}\prod_{v=1}^{(D-3)/2}e^{\frac{i}{2}\,\phi_{2v+1,2v+2}}\,e^{\frac{i}{2}\,(1-\alpha)\left( 1-2q\right) \left( \phi -\frac{t}{\ell}\right) }\,\eta_{s} ~.
\end{equation}
In order for this spinor to be globally well-defined, instead of (\ref{a=1}), the magnetic charge must be quantized:
\begin{equation}
(1-\alpha)\left( 1-2q\right) =n\in \mathbb{Z} ~.
\end{equation}
This is the same condition as in five-dimensional case, Eq.(\ref{quantization}). This leaves us with a BPS configuration preserving $1/2^{\frac{D-1}{2}}$ of the original supercharges.

To conclude, let us note that, in a generic higher-dimensional case, it is sufficient to put an Abelian point-like charge\ (\ref{point-like}) to stabilize a codimension-two brane. The reason why the $D=5$, $N=4$, 2-brane discussed in Sec.\ref{BPS states 5D} needs two Abelian generators taken from a broken $SU(4)$ in order to form a stable configuration, is that this CS theory possesses some \emph{irregular} sectors in the configuration space, that could be treated only if the non-Abelian matter is supported by nonvanishing components $\mathcal{F}$, for $r\neq 0$. Thus, pure-gauge
pont-like charges are not sufficient to provide that. Generic case $N\neq m$ considered here allows, therefore, much more freedom for the choice of matter fields.

%%%%%%%%%%%%%%%%%%%%%%%%%%%%%%%%%%%%%%%%%%%%%%%%%%%%%%%%%%%%%%%%%%%%%%%%%%%%%%%%
\section{Discussion}
%%%%%%%%%%%%%%%%%%%%%%%%%%%%%%%%%%%%%%%%%%%%%%%%%%%%%%%%%%%%%%%%%%%%%%%%%%%%%%%%

In this paper we deal with exact, static, codimension-two brane solutions of Chern-Simons AdS supergravities, coupled to external sources. These solutions, describing local AdS geometries, are systematically constructed in any dimension $D$, using an identification along a Killing vector field with fixed point at the ``center'' of an Euclidean two-dimensional plane, and producing a topological defect of magnitude $\alpha \in (0,1)$. The metric of the resulting $(D-3)$-brane is a naked singularity with mass $M=-(1-\alpha )^2<0$, and becomes a stable BPS state with the addition of an Abelian Chern-Simons charge in such a way that the brane becomes extremal. The extremality condition involves, apart from the mass $M$ and the charge $q$, also some winding number $n\in \mathbb{Z}$. In particular, when the charge is removed ($q=0$), the only supersymmetric solution is the global AdS$_D$, with $\alpha =0$ and $n=-1$. This result is generic, valid in all dimensions.

More explicitly, our analysis starts with the simplest case of three-dimensional supergravity, based on the $OSp(p|2)\times OSp(q|2)$ supergroup, for which we derive the spectrum of BPS 0-branes. This includes a family of supersymmetric extremal naked singularities and also the continuation to positive mass $M\geq 0$ that corresponds to the BTZ black hole.

Next, we move to a more complicated case, that of five-dimensional AdS supergravity for $SU(2,2|N)$ supergroup, and we find the BPS states in a generic sector of the theory where the number of degrees of freedom is maximal. The cases $N=4$ and $N\neq 4$ are treated separately, since the former contains some irregular sectors in the phase space, and its BPS branes possess less supersymmetries. We showed that for $N=4$ these 2-branes saturate the Bogomol'nyi bound. We generalize these results to arbitrary higher odd dimensions.

A point to note in our calculations is that, in solving the Killing equation for the spinor $\epsilon $, one writes a $u(1)$-valued closed 1-form as $b=q\,d\phi_{12}$. Since the Killing equation is local and one wants to be sure the spinor is everywhere well defined, one has to write
$b$ as an exact form everywhere in $\mathbb{R}^2$-$\{0\}$. This can be done defining the angle function $\phi_{12}$ in different patches in such a way that on the intersection of these patches the angles differ by a constant, which is just the period $2\pi(1-\alpha)$. Then, again at such intersection, evaluating the Killing spinor in either patch should give the same result up to a phase due to the $u(1)$ gauge freedom. Finally, since the Killing spinor has a $e^{i\phi_{12}(...)}$ factor, one obtains the same quantization condition as in the previous sections, and the spinor can be periodic or antiperiodic.

Another puzzling issue is that $M<0$ states can be stable even though they look like negative energy states. The answer is that $M$ is a parameter that characterizes the conical defect, $M=-(1-\alpha )^{2}$, while the energy has to be defined as a conserved charge associated to the AdS boost $\mathbf{P}_{0}$ (that, in the $\ell \rightarrow \infty $ limit, becomes a standard Poincar\'e timelike translation). The Bogomol'nyi inequality then ensures that the energy is bounded from below. An example of a spacetime with negative mass is global AdS, that is stable and maximally supersymmetric with $M=-1$.

The $p$-branes described in this paper greatly enrich the spectrum of states of Chern-Simons supergravity. In particular, this has interesting consequences for the case of three dimensions, where standard Einstein-Hilbert gravity and Chern-Simons gravity meet. The negative mass $0$-branes in three dimensional supergravity should contribute to the partition function. It has been conjectured that configurations of this kind might correspond to the missing non-normalizable states required to account for the Bekenstein entropy in the Liouville representation of three dimensional gravity \cite{Krasnov2001,Carlip2005a}. Interestingly enough, this problem was recently revisited in \cite{Maloney2010}, where the authors scrutinize several possible missing contributions to the partition function that may account for the troublesome writing of a physically sensible expression for it. It is natural to wonder whether the $0$-branes discussed in this paper may play a role in that discussion.

BPS and non-BPS $p$-branes in standard supergravity display interesting intersection rules that can be seen as arising from the no-force conditions or compatibility of chirality projections on the Killing spinors (see, for example, \cite{Argurio1997,Ohta1997,Edelstein1998b}). One may wonder about the corresponding statements in the case of $p$-branes coupled to Chern-Simons supergravity. The full answer to this problem is an interesting avenue for future research. In this article, we made the first step in that direction by analyzing the case of intersecting 2-branes in five dimensions in Appendix \ref{intersecting 2-branes in 5D}. We showed that these 2-branes without matter do not intersect in a way compatible with supersymmetry. The question as whether the addition of CS matter can stabilize the branes still remains open.

An interesting generalization to be considered is that of $p$-branes of codimension higher than two. They can be produced by implementing deficit solid angles in the appropriate spheres \cite{Miskovic2009}. The question whether there are BPS configurations of this kind and under which conditions they might exist, should be possible to answer following the approach introduced
in this paper. In the case of codimension-two branes, it is sufficient to have a $U(1)$ gauge field to stabilize the system, that becomes a BPS configuration. A global $U(1)$ can be gauged such that the gravitino becomes charged and the Bohm-Aharonov phase cancels the contribution resulting from the conical defect. This is actually the same mechanism that embodies the twisting procedure in finding supersymmetric wrapped D-branes solutions in standard supergravity \cite{Maldacena2001a,Edelstein2001a}.

We have restricted our discussion to the case of static $2$-branes. A natural step forward is to scrutinize a more general situation in which the angular momentum is prompted into the system, \textit{i.e.}, the case of spinning $p$-branes \cite{Miskovic2009,Edelstein2010a}. The existence of spinning BPS $p$-branes, either with the addition of matter or without it, is a rather interesting problem. In particular, it was shown that singularities arising in BPS solutions of standard supergravity are smoothed out by the inclusion of angular momentum \cite{Maldacena2002}. Whether this phenomenon also takes place in Chern-Simons supergravity is to be discussed in the near future.

%%%%%%%%%%%%%%%%%%%%%%%%%%%%%%%%%%%%%%%%%%%%%%%%%%%%%%%%%%%%%%%%%%%%%%%%%%%%%%%%
\section*{Acknowledgments}
%%%%%%%%%%%%%%%%%%%%%%%%%%%%%%%%%%%%%%%%%%%%%%%%%%%%%%%%%%%%%%%%%%%%%%%%%%%%%%%%

We are pleased to thank Steve Carlip, Anda Degeratu, Marco Farinati, Gast\'{o}n Giribet, Leandro Lombardi, Cristi\'{a}n Mart\'{\i}nez, Ilarion Melnikov, Albert Roura and Ricardo Troncoso for helpful and interesting discussions.
This work is supported in part by MICINN and FEDER (grant
FPA2008-01838), by Xunta de Galicia (Conseller\'\i a de Educaci\'on
and grant PGIDIT06PXIB206185PR), by the Spanish Consolider-Ingenio
2010 Programme CPAN (CSD2007-00042), by a bilateral agreement MINCyT
Argentina (ES/08/02) -- MICINN Spain (FPA2008-05138-E), and by
FONDECYT Grants 11070146,  1100755 and 7080201. A.G. was also
partially supported by UBA-Doctoral Fellowship 572/08. O.M. is also
supported by the PUCV through the projects 123.797/2007,
123.705/2010 and MECESUP Grant UCV0602.
The Centro de Estudios Cient\'\i ficos (CECS) is funded by the Chilean Government
through the Millennium Science Initiative and the Centers of Excellence Base
Financing Program of Conicyt, and by the Conicyt grant ``Southern Theoretical
Physics Laboratory'' ACT-91. CECS is also supported by a group of private companies
which at present includes Antofagasta Minerals, Arauco, Empresas CMPC, Indura,
Naviera Ultragas and Telef\'onica del Sur.

\appendix{}

%%%%%%%%%%%%%%%%%%%%%%%%%%%%%%%%%%%%%%%%%%%%%%%%%%%%%%%%%%%%%%%%%%%%%%%%%%%%%%%%
\section{Conical singularity \label{RegCone}}
%%%%%%%%%%%%%%%%%%%%%%%%%%%%%%%%%%%%%%%%%%%%%%%%%%%%%%%%%%%%%%%%%%%%%%%%%%%%%%%%

A conical singularity is produced in a flat $2D$-plane with polar coordinates $\left( B,\phi _{12}\right) $, $B\geq 0$, identifying the angular coordinate
\begin{equation}
\phi _{12}\simeq \phi _{12}+2\pi (1-\alpha) ~, \qquad 0 < \alpha < 1 ~.
\end{equation}
The angular sector $2\pi \alpha $ is cut out of the plane, the resulting edges being identified, and the surface becomes the cone
\begin{equation}
ds^{2}=dB^{2}+B^{2}d\phi _{12}^{2} ~,
\label{metric_cone}
\end{equation}
with $0 \leq \phi _{12} \leq 2 \pi \left( 1 - \alpha \right)$. The coordinates $\left(B,\phi _{12}\right) $ cover the surface of the cone with the apex at $B=0$, while the coordinates $(x^{1},x^{2})=(r\cos \phi ,r\sin \phi )$ represent the projection of the cone to the $x^{1}$-$x^{2}$-plane, with the apex at $r=0$. The two sets of coordinates are related by
\begin{equation}
B=\frac{r}{1-\alpha} ~,\qquad \phi_{12} = \left( 1-\alpha \right) \phi ~.
\end{equation}
The angle between the generatrix of the cone and its axis is $\theta_{0}=\arcsin \left( 1-\alpha \right) $.

The conical singularity at $r=0$ can be geometrically regularized by defining the apex of the cone as the limit of a spherical cap of vanishing radius. Cutting the cone along the circle $r=r_{0}$, the tip of the cone can be replaced by a spherical cap of radius $\varepsilon$, with $r_{0}=\varepsilon \cos \theta _{0}$. In this way, the two pieces become smoothly joined as a single surface, $\Sigma _{\varepsilon }$. The surface element of this regularized cone is
\begin{equation}
d^{2}s_{\varepsilon} = \left\{
\begin{array}{ll}
\varepsilon^{2} \sin \theta\, d\theta\, d\phi ~,\, & r<r_{0} ~,\quad \theta \in \left[ 0,\frac{\pi }{2}-\theta_{0}\right] ~, \\ [0.8em]
B\, dB\, d\phi _{12} , & r>r_{0} ~.
\end{array} \right.
\end{equation}
The scalar curvature is given by
\begin{equation}
R_{\varepsilon }=\left\{
\begin{array}{ll}
2/\varepsilon ^{2} ~, & r<r_{0} ~, \\ [0.8em]
0 ~, & r>r_{0} ~.
\end{array} \right.
\label{R regularized}
\end{equation}
In the limit $\varepsilon \rightarrow 0$, $R_{\varepsilon}$ must be interpreted as a distribution. Multiplying it by a test function, $\Psi (\vec{r})=\Psi \left( r,\phi \right)$, and integrating over the regularized cone gives
\begin{equation}
\int_{\Sigma _{\varepsilon }}d^{2}s_{\varepsilon }\,R_{\varepsilon}\,\Psi (\vec{r})=4\pi \int_{0}^{\frac{\pi }{2}-\theta_{0}}\!\!\! d\theta\, \sin \theta \,\bar{\Psi}\left( r\right) ~,
\end{equation}
where $\bar{\Psi}\left( r\right) \equiv \frac{1}{2\pi }\int_{0}^{2\pi }d\phi\,\Psi \left( r,\phi \right) $. Since $r=\varepsilon \sin \theta $, we obtain
\begin{equation}
\lim_{\varepsilon \rightarrow 0}\int_{\Sigma_{\varepsilon}}d^{2}s_{\varepsilon }\,R_{\varepsilon }\,\Psi (\vec{r})=4\pi \alpha \,\bar{\Psi}\left( 0\right) ~,
\end{equation}
from which we conclude that the scalar curvature is given by $R=\lim_{\varepsilon \rightarrow 0}R_{\varepsilon }=4\pi \alpha \,\delta (\vec{r})$. In two dimensions, $R$ is uniquely related to the Riemann curvature 2-form as $R^{ab}=\frac{1}{2}\,\varepsilon ^{ab}R\,dx^{1}\wedge dx^{2}$, so that we can write its only nonvanishing component as
\begin{equation}
R^{12}=2\pi \alpha \,\delta (x^{1})\delta (x^{2})\,dx^{1}\wedge dx^{2} ~.
\end{equation}
On the other hand, from $R^{12}=d\omega ^{12}=-dd\phi _{12}$, where $\omega^{ab}$ is the spin connection of the cone calculated from (\ref{metric_cone}), we conclude that geometrical regularization is equivalent to the identity
\begin{equation}
dd\phi_{12} =-2\pi \alpha \,\delta (\Sigma_{12}) ~,
\end{equation}
where $\delta(\Sigma_{12}) \equiv \delta(x^{1})\,\delta(x^{2})\,dx^{1}\wedge dx^{2}$, that is valid for the entire cone.

%%%%%%%%%%%%%%%%%%%%%%%%%%%%%%%%%%%%%%%%%%%%%%%%%%%%%%%%%%%%%%%%%%%%%%%%%%%%%%%%
\section{0-branes in CS supergravity in three dimensions \label{3Dcase}}
%%%%%%%%%%%%%%%%%%%%%%%%%%%%%%%%%%%%%%%%%%%%%%%%%%%%%%%%%%%%%%%%%%%%%%%%%%%%%%%%

%%%%%%%%%%%%%%%%%%%%%%%%%%%%%%%%%%%%%%%%%%%%%%%%%%%%%%%%%%%%%%%%%%%%%%%%%%%%%%%%
\subsection{Super AdS$_{3}$ group \label{Super AdS3}}
%%%%%%%%%%%%%%%%%%%%%%%%%%%%%%%%%%%%%%%%%%%%%%%%%%%%%%%%%%%%%%%%%%%%%%%%%%%%%%%%

The minimal supersymmetric extension of the AdS group in three dimensions is $OSp(p|2)\times OSp(q|2)$, with the generators $\mathbf{G}_{K}=\left\{\mathbf{P}_{a},\mathbf{J}_{ab};\mathbf{T}_{IJ}^{+},\mathbf{T}_{I^{\prime}J^{\prime }}^{-};\mathbf{Q}_{\alpha }^{+I},\mathbf{Q}_{\alpha }^{-I^{\prime
}}\right\}$ \cite{Achucarro1989a}. The $SO(2,2)$ generators in AdS$_{3}$, $\mathbf{P}_{a}$ and $\mathbf{J}_{ab}$ ($a=0,1,2$) can be conveniently redefined as
\begin{equation}
\mathbf{J}_{a}^{\pm }=\frac{1}{2}\,\left( \frac{1}{2}\,\varepsilon_{abc}\,\mathbf{J}^{bc}\pm \mathbf{P}_{a}\right) ~,
\end{equation}
in order to make explicit the isomorphism $SO(2,2)\simeq Sp(2)\times Sp(2)$. The generators $\mathbf{J}_{a}^{+}$ and $\mathbf{J}_{a}^{-}$ commute, and each copy satisfies
\begin{equation}
\left[ \mathbf{J}_{a},\mathbf{J}_{b}\right] = \varepsilon_{abc}\,\mathbf{J}^{c} ~.
\end{equation}
The bosonic sector also contains the internal group $O(p)\times O(q)$ with generators $\mathbf{T}_{IJ}^{+} = - \mathbf{T}_{JI}^{+}$ ($I,J=1,\ldots ,p$) and $\mathbf{T}_{I^{\prime}J^{\prime}}^{-} = - \mathbf{T}_{J^{\prime}I^{\prime}}^{-}$ ($I^{\prime},J^{\prime}=1,\ldots q$), respectively, where each set of rotation generators satisfies
$$
\left[ \mathbf{T}_{IJ},\mathbf{T}_{KL}\right] =\delta_{IL} \mathbf{T}_{JK} - \delta_{JL}\mathbf{T}_{IK} - \delta_{IK}\mathbf{T}_{JL} + \delta_{JK} \mathbf{T}_{IL} ~.
$$
Finally, the supersymmetric generators are $\mathcal{N}=p+q$ real spinors $\mathbf{Q}_{\alpha }^{+I}$ and $\mathbf{Q}_{\alpha }^{-I^{\prime }}$ ($\alpha =1,2$ are spinorial indices) that transform in the vector representation of $O(p)$ and $O(q)$, respectively.

With the above definitions, the generators of the super AdS$_{3}$ algebra split into two commuting sets $\mathbf{G}_{K}=\left\{ \mathbf{G}_{K}^{+}; \mathbf{G}_{K}^{-}\right\}$, where $\mathbf{G}_{K}^{\pm}=\left\{ \mathbf{J}_{a}^{\pm },\mathbf{T}_{IJ}^{\pm},\mathbf{Q}_{\alpha}^{\pm I}\right\}$. Each set of generators satisfies the graded commutator algebra
\begin{eqnarray}
& & \left[ \mathbf{J}_{a},\mathbf{Q}_{\alpha }^{I}\right] = - \frac{1}{2}\,\left( \Gamma _{a}\right) _{\alpha }^{\ \ \beta }\,\mathbf{Q}_{\beta}^{I} ~, \nonumber \\ [0.9em]
& & \left[ \mathbf{T}^{IJ},\mathbf{Q}_{\alpha }^{K}\right] = (\tau^{IJ})^K_{\,\ L}\mathbf{Q}_{\alpha }^{L} ~, \label{algebra} \\ [0.9em]
& & \{\mathbf{Q}_{\alpha }^{I},\mathbf{Q}_{\beta }^{J}\} = c \left[ 2\delta^{IJ}\left( C\,\Gamma ^{a}\right)_{\alpha \beta }\,\mathbf{J}_{a} + C_{\alpha\beta}\,\mathbf{T}^{IJ}\right] ~. \nonumber
\end{eqnarray}
The $\tau$ matrices have components $(\tau^{IJ})^K_{\,\ L} = \delta^I_{\ L}\,\delta^{JK} - \delta^J_{\ L}\,\delta^{IK}$. The conjugation matrix, $C = C_{\alpha\beta} = \varepsilon_{\alpha\beta}$, lowers (and $C^{-1}=C^{\alpha \beta}$ raises) spinor indices ($\varepsilon_{12}=+1$), and the number $c = \pm 1$ distinguishes the two inequivalent representations of $\Gamma$-matrices in three dimensions.

We use a standard representation for the $\Gamma$-matrices and generators of $OSp(p|2)\times OSp(q|2)$ as, for example, the one in Ref.\cite{Giacomini2007}.

%%%%%%%%%%%%%%%%%%%%%%%%%%%%%%%%%%%%%%%%%%%%%%%%%%%%%%%%%%%%%%%%%%%%%%%%%%%%%%%%
\subsection{Absence of Killing spinors for the $0$-brane in minimally supersymmetric AdS$_{3}$
\label{SolKilling3D}}
%%%%%%%%%%%%%%%%%%%%%%%%%%%%%%%%%%%%%%%%%%%%%%%%%%%%%%%%%%%%%%%%%%%%%%%%%%%%%%%%

The ansatz for the metric of a static three-dimensional $0$-brane displayed in (\ref{AdS metric}) can be described by the vielbein $e^{a}$ and the spin connection $\omega^{a} \equiv \frac{1}{2} \varepsilon^a_{\ bc}\, \omega^{bc}$, at $r\neq 0$, as
\begin{eqnarray*}
& & e^{0} = A\,d\phi_{03} ~, \qquad e^{1} = \dfrac{\ell}{A}\,dB ~, \qquad e^{2} = B\,d\phi_{12} ~, \\ [0.9em]
& & \omega^{0} = - \dfrac{A}{\ell}\,d\phi_{12} ~, \qquad \omega^{1} = 0 ~, \qquad \omega^{2} = \dfrac{B}{\ell}\,d\phi_{03} ~,
\end{eqnarray*}
where $A^{2}-B^{2}=\ell^{2}$, $B=r/(1-\alpha)$, $\phi_{12}=(1-\alpha) \phi$, and $\phi_{03} = (1-\alpha)\,t/\ell$. We want to solve the Killing spinor equation (\ref{Killingpm}) for $\boldsymbol{\epsilon}^{+} = \epsilon _{I}^{+\alpha}\,\mathbf{Q}_{\alpha}^{+I} \equiv \epsilon^{+}\mathbf{Q}^{+}$. In Chern-Simons AdS$_{3}$ supergravity, considering the gauge connection $\mathbf{A}$ describing only the 0-brane  ({\it i.e.}, without additional $O(p)\times O(q)$ gauge fields switched on),
\begin{equation}
D\epsilon^{+} = \left[ d - \frac{1}{2}\,\left( \omega^{a} + \frac{1}{\ell}\,e^{a} \right) \Gamma_a \right] \epsilon^{+}=0 ~.
\label{3D Killing}
\end{equation}
Here, $\Gamma_{a}$ are three-dimensional $\Gamma $-matrices and, for simplicity, we choose only one of the two inequivalent representation of $\Gamma$-matrices, with $c=1$. In our notation, $\varepsilon^{012} = 1$.

The radial component of the Killing spinor equation,
\begin{equation}
D_{r}\epsilon^{+} = \left( \partial_{r}-\frac{1}{2 (1-\alpha) A}\,\Gamma_{1}\right) \epsilon^{+}=0 ~,
\end{equation}
has the general solution
\begin{equation}
\epsilon^{+} = e^{f(r) \Gamma _{1}}\,\xi ^{+}(t,\phi) ~,
\end{equation}
where $\xi^{+}$ is a spinor and
\begin{widetext}
\begin{equation}
f(r) = \frac{1}{2}\,\int_{0}^{r/(1-\alpha)}\!\frac{dr^{\prime}}{\sqrt{r^{\prime 2} + (1-\alpha)^{2} \ell^{2}}} = \frac{1}{2}\,\sinh^{-1}\!\left( \frac{r}{(1-\alpha) \ell} \right) ~.
\label{f(r)}
\end{equation}
The two remaining components of the equation are
\begin{eqnarray}
& & \left[ \partial_{\phi} - \frac{1 - \alpha}{2\ell}\, e^{-2f \Gamma_1} (A + B\,\Gamma_1)\,\Gamma_0 \right] \xi^{+} = 0 ~,  \label{3D-12} \\ [0.8em]
& & \left[ \partial_{t} - \frac{1 - \alpha}{2\ell^{2}}\, e^{-2f \Gamma_1} (A + B\,\Gamma_1)\, \Gamma_0 \right] \xi^{+} = 0 ~. \label{3D-03}
\end{eqnarray}
\end{widetext}
It turns out that $f(r)$ satisfies the identity
\begin{equation}
\ell e^{\pm 2f\Gamma_1} = A \pm B\,\Gamma_1 ~,
\end{equation}
and therefore
\begin{equation}
e^{\mp 2f \Gamma_1} \left( A \pm B\,\Gamma_1 \right) = \ell ~.
\label{iden}
\end{equation}
Thus, the general solution of (\ref{3D-12})-(\ref{3D-03}) reads $\xi^{+} = e^{\frac{1}{2}\,(1-\alpha)\,\Gamma_0 \left( \phi +\frac{t}{\ell} \right)}\,\eta^{+}$, and
\begin{equation}
\epsilon^{+} = e^{f(r) \Gamma_1}\, e^{\frac{1}{2}\,(1-\alpha)\,\Gamma_0 \left( \phi + \frac{t}{\ell} \right)}\,\eta^{+} ~.
\label{E_+}
\end{equation}
Here, $\eta ^{+}$ is a constant spinor that can always be chosen as an eigenvector of the matrix $\Gamma_0$, for instance
\begin{equation}
\Gamma_0\,\eta^{+} = i \eta ^{+} ~.
\end{equation}
The Killing spinor $\epsilon^{+}$ has to be globally single-valued, that is, it must be either periodic or antiperiodic under rotations by $2\pi$: $\epsilon ^{+}(\phi + 2\pi) = \pm \epsilon^{+}(\phi)$. This is satisfied by (\ref{E_+}) provided the topological defect is quantized,
\begin{equation}
\alpha = n \in \mathbb{Z} ~.
\end{equation}
Because $\alpha \in \lbrack 0,1)$, one must have $n=0$, that corresponds to global AdS$_{3}$. We conclude that purely gravitational static $0$-branes in three-dimensional $\mathcal{N}=1$ CS supergravity with all additional matter fields switched off do not admit Killing spinors (they are not BPS states).

%%%%%%%%%%%%%%%%%%%%%%%%%%%%%%%%%%%%%%%%%%%%%%%%%%%%%%%%%%%%%%%%%%%%%%%%%%%%%%%%
\section{2-branes in CS supergravity in five dimensions  \label{5D case}}
%%%%%%%%%%%%%%%%%%%%%%%%%%%%%%%%%%%%%%%%%%%%%%%%%%%%%%%%%%%%%%%%%%%%%%%%%%%%%%%%

%%%%%%%%%%%%%%%%%%%%%%%%%%%%%%%%%%%%%%%%%%%%%%%%%%%%%%%%%%%%%%%%%%%%%%%%%%%%%%%%
\subsection{Super AdS$_{5}$ group \label{SAdS5}}
%%%%%%%%%%%%%%%%%%%%%%%%%%%%%%%%%%%%%%%%%%%%%%%%%%%%%%%%%%%%%%%%%%%%%%%%%%%%%%%%

The AdS group in five dimensions is isomorphic to $SU(2,2)$. Its supersymmetric extension is the super unitary group $SU(2,2\left\vert N\right. )$, whose bosonic sector is AdS$_{5}\times SU(N)\times U(1)$. The $su(2,2\left\vert N\right. )$ algebra is spanned by the AdS generators $\mathbf{J}_{AB}=\left( \mathbf{J}_{ab,}\mathbf{P}_{a}\right) $ $\left( A,B=0,\ldots ,5\right) $, the $SU(N)$ generators $\mathbf{T}_{\Lambda}$, $\left( \Lambda =1,\ldots ,N^{2}-1\right) $, an Abelian generator $Z$, and the supersymmetric generators $Q_{s}^{\alpha},\;\bar{Q}_{\alpha }^{s}$. The AdS transformations include Lorentz rotations  $J_{ab}$, and AdS boosts, $P_{a}\equiv J_{a5}\;(a,b=0,\ldots ,4)$, that leave invariant the bilinear form $\eta _{AB}=$  diag $\left( -,+,+,+,+,-\right) $ the fermionic generators are labeled by spinorial index $\alpha=1,\ldots ,4$ and by $s=1,\ldots ,N$.

The dimension of this superalgebra is $N^{2}+8N+15$. For $N=1$, the generators $\mathbf{T}_{\Lambda }$ are absent, and the bosonic sector is given by AdS$_{5}\times u(1)$ algebra. The structure constants and invariant tensor of the superalgebra are calculated in \cite{Chandia1998a,Miskovic2006}, whose notation we follow here. The bosonic generators $J_{AB}$, $T_{\Lambda }$  and $Z$ form the algebra $su(2,2)\times su(N)\times u(1)$, while the supersymmetry generators transforms as spinors under AdS and as vectors
under $su(N)$,
\begin{eqnarray}
& \left[ \mathbf{J}_{AB},\mathbf{Q}_{s}^{\alpha }\right] = - \frac{1}{2}\,\left( \Gamma_{AB} \right) _{\beta }^{\alpha }\,\mathbf{Q}_{s}^{\beta } ~, \qquad & \left[ \mathbf{T}_{\Lambda},\mathbf{Q}_{s}^{\alpha }\right] = \left( \tau _{\Lambda }\right) _{s}^{r}\,\mathbf{Q}_{r}^{\alpha} ~, \nonumber \\ [0.8em]
& \left[ \mathbf{J}_{AB},\mathbf{\bar{Q}}_{\alpha }^{s}\right] =\frac{1}{2}\,\mathbf{\bar{Q}}_{\beta}^{s}\,\left( \Gamma _{AB}\right) _{\alpha }^{\beta} ~, \qquad & \left[ \mathbf{T}_{\Lambda },\mathbf{\bar{Q}}_{\alpha }^{s}\right] = - \mathbf{\bar{Q}}_{\alpha }^{r}\,\left( \tau _{\Lambda }\right) _{r}^{s} ~, \nonumber
\end{eqnarray}
where the $4\times 4$ matrices $\Gamma _{AB}$ are defined as $\Gamma _{ab}=\frac{1}{2}\,\left[ \Gamma _{a},\Gamma _{b}\right] $ and $\Gamma_{a5}=\Gamma _{a}$, and the Dirac matrices in five dimensions $\Gamma _{a}$ satisfy $\Gamma _{0}^{\dag }=-\Gamma _{0}$ and $\Gamma _{a}^{\dag }=\Gamma_{a}$ for $a\neq 0$. Also, $\tau _{\Lambda }$ are anti-Hermitean $N\times N$ matrices, generators of $su(N)$. The SUSY generators carry $u(1)$ charges,
\begin{equation}
\left[ \mathbf{Z},\mathbf{Q}_{s}^{\alpha }\right] =-i\,\left( \frac{1}{4} - \frac{1}{N}\right) \,\mathbf{Q}_{s}^{\alpha } ~,\qquad \left[ \mathbf{Z}, \mathbf{\bar{Q}}_{\alpha }^{s}\right] =i\,\left( \frac{1}{4}-\frac{1}{N} \right) \,\mathbf{\bar{Q}}_{\alpha }^{s} ~,
\label{U(1) Charges}
\end{equation}
and the anticommutator of the supersymmetry generators has the form
\begin{equation}
\left\{ \mathbf{Q}_{s}^{\alpha }\mathbf{,\bar{Q}}_{\beta }^{r}\right\} = \frac{1}{4}\,\delta _{s}^{r}\,\left( \Gamma ^{AB}\right) _{\beta }^{\alpha}\,\mathbf{J}_{AB}-\delta _{\beta }^{\alpha }\,\left( \tau ^{\Lambda}\right) _{s}^{r}\,\mathbf{T}_{\Lambda }+i\,\delta _{\beta }^{\alpha
}\,\delta _{s}^{r}\,\mathbf{Z} ~.
\end{equation}
For $N=4$, the Abelian generator $\mathbf{Z}$ becomes a central charge and the superalgebra becomes a central extension of $PSU(2,2\left\vert 4\right. )$.

An invariant third rank tensor, completely symmetric in bosonic and antisymmetric in fermionic indices, can be constructed as
\begin{equation}
i\,g_{KLM}\equiv \left\langle \mathbf{G}_{K}\mathbf{G}_{L}\mathbf{G}_{M}\right\rangle =\frac{1}{2}\,\text{Str\thinspace }\left[ \left( \mathbf{G}_{K}\mathbf{G}_{L}+\left( -\right) ^{\varepsilon _{K}\varepsilon _{L}} \mathbf{G}_{L}\mathbf{G}_{K}\right) \mathbf{G}_{M}\right] ~,
\label{3}
\end{equation}
and it has the following nonvanishing components,
\begin{equation}
\begin{array}{ll}
g_{[AB] [CD] [EF]} = -\displaystyle\frac{1}{2}\,\varepsilon_{ABCDEF} ~, \qquad & \qquad g_{Z [AB] [CD]} = -\displaystyle\frac{1}{4}\,\eta_{[AB] [CD]} ~, \\ [1em]
g_{\Lambda_1 \Lambda_2 \Lambda_3} = -\gamma_{\Lambda_1 \Lambda_2 \Lambda_3} ~, \qquad & \qquad g_{Z \Lambda_1 \Lambda_2} = -\displaystyle\frac{1}{N}\,\gamma_{\Lambda_1 \Lambda_2} ~, \\ [1em]
g_{[AB] {\alpha \choose r} {s \choose \beta}} = -\displaystyle\frac{i}{4}\,\left( \Gamma_{AB} \right)_{\beta}^{\alpha}\,\delta_{r}^{s} ~, \qquad & \qquad g_{Z {\alpha \choose r} {s \choose \beta}} = \displaystyle\frac{1}{2}\,\left( \frac{1}{4} + \frac{1}{N} \right) \delta_{\beta}^{\alpha}\ \delta_{r}^{s} ~, \\ [1em]
g_{\Lambda {\alpha \choose r} {s \choose \beta}} = -\displaystyle\frac{i}{2}\,\delta_{\beta}^{\alpha} \left( \tau_{\Lambda} \right)_{r}^{s} ~, & \qquad g_{ZZZ} = \displaystyle\frac{1}{N^{2}} - \frac{1}{4^{2}} ~,
\end{array}
\end{equation}
Here $\eta _{\left[ AB\right] \left[ CD\right] }\equiv \eta_{AC}\,\eta _{BD}-\eta _{AD}\,\eta _{BC}\,$ and $\gamma _{\bar{K}\bar{L}}$ are the Killing metrics of $SO(2,4)$ and $SU(N)$, respectively. The symmetric third rank invariant tensor for $su(N)$ is $\gamma _{\Lambda_{1}\Lambda _{2}\Lambda _{3}}\equiv \frac{1}{2i}\,$Tr$_{N}\left( \left\{ \tau _{\Lambda _{1}},\tau _{\Lambda _{2}}\right\} \tau _{\Lambda_{3}}\right) $, and the $\Gamma $-matrices are normalized so that
\begin{equation}
\text{Tr}_{4}\left( \Gamma _{a}\,\Gamma _{b}\,\Gamma _{c}\,\Gamma_{d}\,\Gamma _{e}\right) =-4i\,\varepsilon _{abcde} ~,
\end{equation}
$(\varepsilon ^{abcde5}:=\varepsilon ^{abcde},~\varepsilon ^{012345}=1)$, that is consistent with $\Gamma _{0}=i\Gamma _{1}\Gamma _{2}\Gamma_{3}\Gamma _{4}$.

Splitting the generators as $\mathbf{G}_{K}=\left( \mathbf{G}_{\bar{K}}, \mathbf{Z}\right) $, the invariant tensor for $SU(2,2|N)$ has $g_{\bar{K} \bar{L}Z}$ invertible and $g_{\bar{K}ZZ}=0$. In the special case $N=4$, the invariant tensor $g_{KLM}$ of $SU(2,2\left\vert 4\right. )$ simplifies to $g_{z\bar{K}\bar{L}}=-\frac{1}{4}\,\gamma _{\bar{K}\bar{L}}$ and $g_{ZZZ}=0$, where $\gamma _{\bar{K}\bar{L}}$ is the Killing metric for $PSU(2,2|4)$.

%%%%%%%%%%%%%%%%%%%%%%%%%%%%%%%%%%%%%%%%%%%%%%%%%%%%%%%%%%%%%%%%%%%%%%%%%%%%%%%%
\subsection{Killing spinors without additional matter  \label{pure 5D 2-brane}}
%%%%%%%%%%%%%%%%%%%%%%%%%%%%%%%%%%%%%%%%%%%%%%%%%%%%%%%%%%%%%%%%%%%%%%%%%%%%%%%%

In this section we integrate the Killing spinor equation for a single static 2-brane in five dimensions,
\begin{equation}
D\epsilon =\left( d+\frac{1}{4}\,\omega ^{ab}{\Gamma }_{ab}+\frac{1}{2\ell }%
\,e^{a}{\Gamma }_{a}\right) \epsilon = 0 ~,
\label{Psi_eq}
\end{equation}
where the vielbein of a 2-brane is
\begin{equation}
\begin{array}{l}
e^{0} = A\,\cosh\rho\,d\phi_{05} ~, \qquad e^{1} = \dfrac{\ell}{A}\,dB ~, \qquad e^{2} = B\,d\phi_{12} ~, \\ [1em]
e^{3} = A\,d\rho ~, \qquad e^{4} = A\,\sinh\rho\,d\theta ~.
\end{array}
\label{2brane_e}
\end{equation}
As before, $A^{2}-B^{2}=\ell ^{2}$. Assuming a torsionless spacetime, $T^{a}=0$\thinspace , which is true  everywhere outside the singularity, the nonvanishing components of the spin connection are found to be
\begin{equation}
\begin{array}{ll}
\omega^{01}=\dfrac{B}{\ell }\,\cosh \rho \,d\phi_{05} ~, & \ \omega ^{13} = - \dfrac{B}{\ell }\,d\rho ~, \\ [1em]
\omega^{03}=\sinh \rho \,d\phi _{05} ~, & \ \omega ^{14}=-\dfrac{B}{\ell }%
\,\sinh \rho \,d\theta ~, \\ [1em]
\omega^{12}=-\dfrac{A}{\ell }\,d\phi_{12} ~, & \ \omega ^{34}=-\cosh \rho
\,d\theta ~.
\end{array}
\label{2brane_w}
\end{equation}
Again, rescaling the coordinates as $B=r/(1-\alpha)$, $\phi_{12}=(1-\alpha )\phi $  and $\phi_{05}=(1-\alpha )t/\ell$, the radial component of Eq.(\ref{Psi_eq}) has the form
\begin{equation}
\left( \partial _{r}+\frac{1}{2 (1-\alpha) A}\,\Gamma _{1}\right) \epsilon = 0 ~,
\end{equation}
and the solution is
\begin{equation}
\epsilon =e^{-f\Gamma _{1}}\chi (t,\phi ,\rho ,\theta) ~,
\label{psi}
\end{equation}
and the radial function $f(r)=\frac{1}{2}\sinh ^{-1}\left( \frac{r}{(1-\alpha) \ell} \right) $ has the same form as in three dimensions, Eq.(\ref{f(r)}). The equation along $\rho $, $e^{f\Gamma _{1}}D_{\rho }\epsilon =0$, then becomes
\begin{equation}
\left( \partial _{\rho }+\frac{1}{2\ell }\,e^{2f\Gamma_{1}}\left( A-B\Gamma_{1} \right) \Gamma_{3}\right) \chi = 0 ~.
\label{taueq2}
\end{equation}

Even though $A(r)$ and $B(r)$ are functions of $r$, the radial dependence in this equation drops out thanks to the identity (\ref{iden}), becoming just $\left( \partial _{\rho }+\frac{1}{2}\,\Gamma _{3}\right) \chi=0$, whose solution is
\begin{equation}
\chi =e^{-\frac{1}{2}\,\rho \Gamma _{3}}\lambda (t,\phi ,\theta) ~.
\end{equation}
The component of the Killing spinor equation along $\phi $ is similarly simplified
\begin{equation}
\left( \partial _{\phi }-\frac{1-\alpha}{2}\,\Gamma _{12}\right) \lambda = 0 ~,
\end{equation}
whose integral is
\begin{equation}
\lambda =e^{\frac{1}{2}\,(1-\alpha)\phi \Gamma _{12}}\xi (t,\theta ) ~.
\end{equation}
The component along $t$ becomes $\left( \partial _{t}+\frac{a}{2\ell }\,\Gamma _{0}\right) \xi =0$ , and integrates to
\begin{equation}
\xi =e^{-\frac{(1-\alpha)t}{2\ell }\,\Gamma _{0}}\varphi (\theta) ~.
\end{equation}
The last component corresponding to the coordinate $\theta$ is solved in anbanalogous manner with $\varphi (\theta )=e^{\frac{1}{2}\,\theta \Gamma_{34}}\eta $, so the final form of the Killing spinor is
\begin{equation}
\epsilon =e^{-f\Gamma _{1}}e^{-\frac{1}{2}\,\rho \Gamma _{3}}e^{\frac{1}{2}\,(1-\alpha)\phi \Gamma _{12}}e^{-\frac{(1-\alpha)t}{2\ell }\Gamma _{0}}e^{\frac{1}{2}\,\theta \Gamma_{34}}\,\eta ~.
\end{equation}
The constant spinor $\eta $ can be chosen as a common eigenvector of two commuting Dirac matrices,
\begin{equation}
\Gamma _{12}\eta =-i\eta ~, \qquad \Gamma _{34}\eta =-i\eta ~,
\end{equation}
For $\epsilon $ to be globally well-defined, with periodic or antiperiodic boundary conditions in the angular coordinates $\phi $ and $\theta $ with the periods $2\pi $, we have to impose the condition
\begin{equation}
\alpha =n\in \mathbb{Z} ~.
\end{equation}
Similarly as in three dimensions, the only allowed solution is $\alpha =0$, that is the global AdS$_{5}$. Thus, static 2-branes in five dimensions without additional fields switched on do not admit Killing spinors.

%%%%%%%%%%%%%%%%%%%%%%%%%%%%%%%%%%%%%%%%%%%%%%%%%%%%%%%%%%%%%%%%%%%%%%%%%%%%%%%%
\subsection{Asymptotic Killing vectors \label{CovConstant}}
%%%%%%%%%%%%%%%%%%%%%%%%%%%%%%%%%%%%%%%%%%%%%%%%%%%%%%%%%%%%%%%%%%%%%%%%%%%%%%%%

We look for the asymptotic Killing vectors $\mathbf{\bar{\lambda}} = \bar{\lambda}^{K}\mathbf{G}_{K}$, solutions of the equation $\bar{D}\mathbf{\bar{\lambda}}=0$ at the spatial boundary $\Sigma_{\infty}$, for the BPS background obtained in Sec. \ref{BPS states 5D}. We assume only the bosonic sector of super AdS$_{5}$ connection to be switched on,
\begin{equation}
\mathbf{\bar{A}}=\mathbf{\bar{A}}_{\text{AdS}}+\left( (1-\alpha)\, q\, \mathbf{T}_{12} + \mathcal{E}r\,\mathbf{Z}\right)\,d\phi +\mathcal{B}\rho\, \mathbf{Z}\,d\theta ~,
\end{equation}
where $(1-\alpha )(q-1)=n\in \mathbb{Z}$, and the spacetime geometry is encoded in the AdS connection for the 2-brane,
\begin{eqnarray}
\mathbf{\bar{A}}_{\text{AdS}} &=&\frac{r}{\ell }\, \left( \mathbf{J}_{25}-\mathbf{J}_{12}\right) \, d\phi + \frac{r}{(1-\alpha )\ell }\,\left( \mathbf{J}_{35}
-\mathbf{J}_{13}\right) \,d\rho  \nonumber \\ [0.9em]
&& +\left[ \frac{r\sinh \rho }{(1-\alpha )\ell }\,\left( \mathbf{J}_{45}- \mathbf{J}_{14}\right) -\cosh \rho\,\mathbf{J}_{34}\right] \,d\theta ~.
\end{eqnarray}
The radial coordinate $r$ on $\Sigma _{\infty }$ is a large fixed parameter. The topology of $\Sigma_{\infty}$ is $H^{2}(\rho ,\theta )\times S^{1}(\phi )$, where $\phi $ and $\theta $ are periodic coordinates, and therefore the solutions must be periodic in $\phi ,\theta $ as well. We will assume that $\alpha \neq 0$ (this is a true 2-brane, and not global AdS).

The equation
\begin{equation}
d\mathbf{\bar{\lambda}}+\left[ \mathbf{\bar{A}}_{\text{AdS}}, \mathbf{\bar{\lambda}}\right] +(1-\alpha +n)\,d\phi\,\left[ \mathbf{T}_{12}, \mathbf{\bar{\lambda}}\right] = 0 ~,
\end{equation}
gives that the Abelian Killing vector ($d\bar{\lambda}^{Z}=0$) is constant,
\begin{equation}
U(1):\qquad \bar{\lambda}^{Z} = const ~,
\end{equation}
and this is a also a symmetry of the solution in the bulk manifold.

For the $SU(4)$ components, we have
\begin{equation}
\bar{D}_{SU(4)} \bar{\lambda}^{IJ} = d \bar{\lambda}^{IJ} + (1-\alpha +n)\,d\phi\, \left( \delta_{2}^{[I}\bar{\lambda}^{J]1} -\delta_{1}^{[I}\bar{\lambda}^{J]2}\right) = 0 ~,
\end{equation}
that becomes $d\bar{\lambda}^{12}=0$ and $d\left. \bar{\lambda}^{IJ}\right\vert _{I,J\in \{3,4,5,6\}}=0$, leading to the asymptotic Killing vectors
\begin{equation}
SU(4):\qquad \bar{\lambda}^{12}\;,\,\left. \bar{\lambda}^{IJ} \right\vert_{I,J\in \{3,4,5,6\}} = const ~,
\label{SU(4)as}
\end{equation}
corresponding to the $U(1)\times SO(4)$ symmetry. Furthermore, taking into account that the functions $\sin (1-\alpha +n)\phi $ and $\cos (1-\alpha+n)\phi $ are not periodic in $\phi $ for $\alpha \in (0,1)$, there are no additional solutions. The non-Abelian asymptotic symmetry (\ref{SU(4)as}) is much larger than the one in the bulk, described by $\bar{\lambda}^{12}$, $\bar{\lambda}^{34}$ and $\bar{\lambda}^{56}$, and reflecting the $U(1)\times U(1)\times U(1)$ invariance of the background.

The AdS components of the asymptotic Killing equation
\[
\bar{D}_{\text{AdS}}\,\bar{\lambda}^{AB}=d\bar{\lambda}^{AB} +\bar{A}_{\text{AdS}}^{AC}\bar{\lambda}_{C}^{\ \ B}
-\bar{A}_{\text{AdS}}^{BC}\bar{\lambda}_{C}^{\ \ A}=0 ~,
\]
have a 1-parameter's solution
\begin{equation}
\text{AdS}_{5}:\qquad \bar{\lambda}^{25} = -\bar{\lambda}^{12} = const ~,
\end{equation}
that corresponds to the $\partial_{\phi}$ isometry.

%%%%%%%%%%%%%%%%%%%%%%%%%%%%%%%%%%%%%%%%%%%%%%%%%%%%%%%%%%%%%%%%%%%%%%%%%%%%%%%%
\subsection{Mode expansion and Bogomol'nyi bound in 5D \label{BPS bound}}
%%%%%%%%%%%%%%%%%%%%%%%%%%%%%%%%%%%%%%%%%%%%%%%%%%%%%%%%%%%%%%%%%%%%%%%%%%%%%%%%

Here we show explicitly that the algebra of conserved charges of 2-branes implies a lower bound for $E$.

We start from the mode expansion of the algebra of supercharges on the spatial boundary isomorphic to $H^{2}\times S^{1}$, parameterized by the coordinates $\rho \in \left[ -\frac{L}{2},\frac{L}{2}\right] $ (where $L\rightarrow \infty $) and $\theta \in \lbrack 0,2\pi ]$, and the circle $S^{1}$ is parameterized by the periodic angle $\phi $. All quantities like the charges can be expanded in Fourier modes of the spatial section as
\begin{equation}
X(r,\rho ,\theta ,\phi )=\int dw\,\sum\limits_{m,k}X_{wmk}(r)\,e^{\frac{2\pi i}{L}\,w\rho +im\theta + ik\phi} ~,
\end{equation}
where the Fourier coefficients are
\begin{equation}
X_{wmk}(r)=\int \frac{d\rho d\theta d\phi }{L\left( 2\pi \right) ^{2}}\,X(r,\rho ,\theta ,\phi )\,e^{-\frac{2\pi i}{L}\,w\rho -im\theta -ik\phi} ~.
\end{equation}
The notation is simplified calling $\vec{s}=(w,m,k)$, and consequently $\sum\limits_{\vec{s}}=\int dw\sum\limits_{m,k}$, $\delta _{\vec{s}, \vec{s}^{\prime}}=\delta (w-w^{\prime })\delta _{mm^{\prime }}\delta _{kk^{\prime}}$, etc. Then, the mode expansions for the canonical and central charges (\ref{Q evaluated}) and (\ref{C explicitly}) read
\begin{eqnarray}
&& Q\left[ \lambda \right] =\sum_{\vec{s}}\lambda_{\vec{s}}^{K}\,q_{K,-\vec{s}} ~,\quad q_{K,\vec{s}}=\frac{\kappa \mathcal{B}}{4}\,\gamma _{KL}\,A_{\phi ,\vec{s}}^{L} ~, \qquad \\ [0.9em]
&& C\left[ \lambda ,\eta \right] = \frac{i\kappa \mathcal{B}}{4}\,\gamma_{KL}\sum\limits_{\vec{s},\vec{s}^{\prime }}\lambda _{\vec{s}}^{K}\,\eta_{\vec{s}^{\prime }}^{L}\,k\,\delta _{\vec{s}+\vec{s}^{\prime },0} ~,
\end{eqnarray}
and the algebra of supersymmetric charges written in modes adopts the form
\begin{equation}
\left\{ q_{K,\vec{s}},q_{L,\vec{s}^{\prime }}\right\} =f_{KL}^{\quad \;M}q_{M,\vec{s} + \vec{s}^{\prime}}+\frac{i\kappa \mathcal{B}}{4}\,k\,\gamma _{KL}\,\delta _{\vec{s} + \vec{s}^{\prime },0} ~.
\label{mode algebra2}
\end{equation}

This is a supersymmetric extension of the WZW$_{4}$ algebra. It has a nontrivial central extension for $psu(2,2\left\vert 4\right. )$ which depends only on the $u(1)$ flux determined by $\mathcal{B}$. The modes $q_{K,\vec{s}}$ with $\vec{s}=(0,0,k)$ form a Kac-Moody subalgebra with the central charge $\kappa \mathcal{B}/4$, while the modes with $\vec{s}=(w,0,0)$ and $(0,m,0)$ form Kac-Moody subalgebras without central charges. For the supersymmetry charges, using $q_{Z}=0$, the algebra reads
$$
\left\{ q_{r,\vec{s}}^{\alpha },\bar{q}_{\beta ,\vec{s}^{\prime }}^{s}\right\} = - \frac{1}{2}\,\delta_{r}^{s}\left( \Gamma ^{a}\right) _{\beta }^{\alpha }\,q_{a,\vec{s}+\vec{s}^{\prime }}+\frac{1}{4}\,\delta_{r}^{s}\left( \Gamma ^{ab}\right) _{\beta }^{\alpha }\,q_{ab,\vec{s}+\vec{s}%
^{\prime}} - \frac{1}{4}\,\delta _{\beta }^{\alpha }\,\left( \hat{\Gamma}^{IJ}\right)
_{r}^{s}\,q_{IJ,\vec{s}+\vec{s}^{\prime }}-\frac{i\kappa \mathcal{B}}{4}\,k\,\delta _{r}^{s}\,\delta _{\beta }^{\alpha }\,\delta _{\vec{s}+\vec{s}^{\prime },\vec{0}} ~.
$$
Multiplying this algebra by $\Gamma ^{0}$ and using $\bar{q}=q^{\dagger }\Gamma _{0}$ (where $\left( \Gamma^{0}\right) ^{2}=-1$), we can construct a semipositive definite matrix $\left\{ q_{r,\vec{0}}^{\alpha },q^{\dagger}{}_{\beta ,\vec{0}}^{s}\right\} $, that leads to the bound
\begin{equation}
- \frac{1}{2}\,\delta _{r}^{s}\left( \Gamma ^{a}\Gamma ^{0}\right) _{\beta }^{\alpha}\,q_{a,\vec{0}}+\frac{1}{4}\,\delta _{r}^{s}\left( \Gamma^{ab}\Gamma ^{0}\right) _{\beta }^{\alpha }\,q_{ab,\vec{0}} -\frac{1}{4}\,\left( \Gamma ^{0}\right) _{\beta }^{\alpha }\,\left( \hat{\Gamma}^{IJ}\right) _{r}^{s}\,q_{IJ,\vec{0}}-\frac{i\kappa \mathcal{B}}{4}\,k\,\left( \Gamma ^{0}\right) _{\beta }^{\alpha }\,\delta _{r}^{s} \geq 0 ~.
\label{Bound}
\end{equation}
Identifying the energy $E=q_{0,\vec{0}}$ with the time component of the AdS boost charge $q_{a,\vec{0}}=(E,q_{\bar{a},\vec{0}})$, where the Lorentz index is decomposed as $a=\left( 0,\bar{a}\right) $, and $\eta _{\bar{a}\bar{b}}=\delta _{\bar{a}\bar{b}}$ ($\bar{a},\bar{b}=1,2,3,4$) is the Euclidean metric, one finds
\[
-\frac{1}{2}\,\delta _{r}^{s}\left( \Gamma ^{a}\Gamma ^{0}\right) _{\beta }^{\alpha}\,q_{a,\vec{0}}=\frac{1}{2}\,\delta _{r}^{s}\delta _{\beta }^{\alpha }\,E-\frac{1}{2}\,\delta _{r}^{s}\left( \Gamma ^{\bar{a}}\Gamma ^{0}\right) _{\beta }^{\alpha }\,q_{\bar{a},\vec{0}} ~,
\]
and the bound (\ref{Bound}) can be rewritten as
\begin{equation}
\delta _{r}^{s}\delta _{\beta }^{\alpha }\,E\geq M_{r\beta }^{s\alpha } ~,
\label{inequality}
\end{equation}
where $\mathbf{M}=[M_{r\beta }^{s\alpha}]$ is an auxiliary matrix,
\begin{equation}
\mathbf{M} \equiv 1\otimes \left( \Gamma ^{\bar{a}}\Gamma ^{0}\,q_{\bar{a},\vec{0}}-\frac{1}{2}\,\Gamma ^{ab}\Gamma ^{0}q_{ab,\vec{0}}\right) +\frac{1}{2}\,\hat{\Gamma}^{IJ}\,q_{IJ,\vec{0}}\,\otimes \Gamma ^{0}+\frac{i\kappa \mathcal{B}}{4}\,k\,\left( 1\otimes \Gamma ^{0}\right) ~,
\end{equation}
in the basis $v^{r}\otimes \eta _{\alpha }$. Thus, from (\ref{inequality}) it follows that the energy must be larger than all eigenvalues $\lambda _{i}$ of the matrix $\mathbf{M}$,
\begin{equation}
E\geq \lambda _{i} ~, \qquad (\forall i) ~.
\end{equation}
Since the trace of $\mathbf{M}$ vanishes, (\ref{inequality}) implies $E\geq 0 $, the largest $\lambda $ must be nonegative, and it is sufficient to find the eigenvalues of the simpler matrix with the same eigenvalues as $\mathbf{M}$,
\begin{equation}
i\mathbf{M}\Gamma ^{0} = 1\otimes \left( -i\Gamma ^{\bar{a}}q_{\bar{a},\vec{0}}+\frac{i}{2}\,\Gamma ^{ab}q_{ab,\vec{0}}\right) - \frac{i}{2}\,q_{IJ,\vec{0}}\,\hat{\Gamma}^{IJ}\otimes 1+\frac{\kappa \mathcal{B}}{4}\,k\,\left( 1\otimes 1\right) ~.
\end{equation}
For simplicity, we choose the rest frame ($q_{\bar{a},\vec{0}}=0$) and we find that $\left( \frac{i}{2}\Gamma^{ab}q_{ab,\vec{0}}\right) _{\beta }^{\alpha }$ has four eigenvalues $\left\{ \pm \eta_{+},\pm \eta_{-}\right\} $ (with $\eta _{\pm }\geq 0$) given by
$$
\eta_{\pm} = \left[ -\sum_{\bar{a}}q_{0\bar{a},\vec{0}}^{2}+\sum_{\bar{a} < \bar{b}}q_{\bar{a}\bar{b},\vec{0}}^{2} \pm \left( -\left( \varepsilon ^{0bcde}q_{bc,\vec{0}}q_{de,\vec{0}}\right) ^{2}+\sum_{\bar{a}}\left( \varepsilon ^{\bar{a}bcde}q_{bc,\vec{0}}q_{de,\vec{0}}\right) ^{2}\right) ^{1/2}\right] ^{1/2} ~.
$$
Similarly, the matrix $\left( -\frac{i}{2}\,q_{IJ,\vec{0}}\,\hat{\Gamma}^{IJ}\right) _{s}^{r}$ has four eigenvalues $\left\{ \pm v_{+},\pm v_{-}\right\} $, where $v_{\pm }\geq 0$ and
\[
v_{\pm }=\left[ \sum_{I<J}q_{IJ,\vec{0}}^{2}\pm \left( \sum_{I}\left( \varepsilon ^{IJKLM}q_{JK,\vec{0}}q_{LM,\vec{0}}\right) ^{2}\right) ^{1/2} \right] ^{1/2} ~.
\]
The sixteen eigenvalues of the $16\times 16$ matrix $\mathbf{M}$ are then $\lambda =\pm v_{\pm }\pm \eta _{\pm }+\frac{\kappa \mathcal{B}}{4}\,k$, with $2^{4}=16$ independent combinations of $\pm $. Clearly, $E$ is larger than the largest $\lambda $, that finally leads to the bound given by Eq.(\ref{E}),
\begin{equation}
E \geq \max \left\{ v_{\pm }+\eta _{\pm }+\frac{\kappa }{4}\,\left\vert \mathcal{B}\,k\right\vert \right\} ~.
\end{equation}
%

%%%%%%%%%%%%%%%%%%%%%%%%%%%%%%%%%%%%%%%%%%%%%%%%%%%%%%%%%%%%%%%%%%%%%%%%%%%%%%%%
\subsection{Intersecting 2-branes \label{intersecting 2-branes in 5D}}
%%%%%%%%%%%%%%%%%%%%%%%%%%%%%%%%%%%%%%%%%%%%%%%%%%%%%%%%%%%%%%%%%%%%%%%%%%%%%%%%

Here we show that the intersection of two 2-branes in 5 dimensions does not make a BPS 0-brane.

Based on the previous construction for the 2-brane, we take a point-like singularity in AdS$_{5}$ placed at the origin of the $x^{1}$-$x^{2}$ plane, $\Sigma _{12}$, and another point-like singularity in the origin of the orthogonal $x^{3}$-$x^{4}$ plane, $\Sigma_{34}$. The brane then has support at the intersection of the origins of these planes, that has a 1-dimensional worldvolume. In the embedding space, the coordinates are
\begin{equation}
\begin{array}{lll}
x^{0}=A\cos \phi_{05} ~, \qquad & \qquad x^{1}=B\cos \phi_{12} ~,\qquad & \qquad x^{3}=C\cos \phi_{34} ~, \\ [1em]
x^{5}=A\sin \phi_{05} ~, \qquad & \qquad x^{2}=B\sin \phi_{12} ~, \qquad & \qquad x^{4}=C\sin \phi_{34} ~,
\end{array}
\end{equation}
with the AdS constraint $-A^{2}+B^{2}+C^{2}=-\ell ^{2}$ and the angles $\phi _{12}$ and $\phi _{34}$ that have some angular deficits. The metric of this spacetime of constant curvature is
$$
ds^{2} = \left( 1-\frac{B^{2}}{A^{2}}\right) dB^{2}+\left( 1-\frac{C^{2}}{A^{2}}\right) dC^{2}  -2BC\,dB\,dC+B^{2}d\phi _{12}^{2}+C^{2}d\phi _{34}^{2}-A^{2}d\phi _{05}^{2} ~.
$$
The vielbein can be chosen as
\begin{eqnarray}
e^{0} &=&A\,d\phi_{05} ~, \qquad e^{1}=\dfrac{\ell }{\sqrt{\ell ^{2}+B^{2}}}\,dB ~, \qquad e^{2} =B\,d\phi_{12} ~, \nonumber \\ [1em]
e^{3} &=&\dfrac{\sqrt{\ell ^{2}+B^{2}}}{A}\,dC-\dfrac{BC}{A\sqrt{\ell ^{2}+B^{2}}}\,dB ~, \qquad e^{4}=C\,d\phi _{34} ~, \nonumber
\end{eqnarray}
while the nonvanishing components of the Levi-Civit\`{a} spin connection in the space surrounding the singularities read
\begin{eqnarray}
\omega^{01} &=&\dfrac{AB}{\ell \sqrt{\ell ^{2}+B^{2}}}\,d\phi _{05} ~, \qquad \omega^{03} =\dfrac{C}{\sqrt{\ell ^{2}+B^{2}}}\,d\phi _{05} ~, \nonumber \\ [1em]
\omega^{12} &=&-\dfrac{\sqrt{\ell ^{2}+B^{2}}}{\ell }\,d\phi _{12} ~, \qquad \omega^{13} = -\dfrac{B}{\ell A}\left( dC-\dfrac{BC}{\ell ^{2}+B^{2}}\,dB\right) , \\ [1em]
\omega^{14} &=&-\dfrac{BC}{\ell \sqrt{\ell ^{2}+B^{2}}}\,d\phi _{34} ~, \qquad \omega^{34} = -\dfrac{A}{\sqrt{\ell ^{2}+B^{2}}}\,d\phi _{34}~.  \nonumber
\end{eqnarray}

The 0-brane is produced by two independent rotations in the planes $(x^{1},x^{2})$ and $(x^{3},x^{4})$, so we shall assume that both angles $\phi _{12}\in \lbrack 0,2\pi (1-\alpha ))$ and $\phi _{34}\in \lbrack 0,2\pi (1-\beta ))$\emph{\ } have angular deficits with $\alpha ,\beta \in \lbrack 0,1).$ Then we can use the identities $dd\phi _{12}=-2\pi \alpha \delta (\Sigma _{12})$ and $dd\phi _{34}=-2\pi \beta \delta (\Sigma _{34})$ where, as usual, $\delta (\Sigma _{AB})=\delta (x^{A})\delta (x^{B})\,dx^{A}\wedge dx^{B}$. The AdS curvature on the whole manifold has the form
\begin{equation}
\mathbf{F}=2\pi \alpha \,\delta (\Sigma _{12})\,\mathbf{J}_{12}+2\pi \beta \,\delta (\Sigma_{34})\mathbf{J}_{34} ~,
\label{0branecurvature}
\end{equation}
and it is a solution of the Chern-Simons equations of motion
\begin{equation}
\left\langle \left( \mathbf{F}^{2}-\mathbf{j}_{[0]}\right) \mathbf{J}_{AB}\right\rangle = 0 ~,
\end{equation}
where the external current 4-form that defines this 0-brane is
\begin{equation}
\mathbf{j}_{[0]}=8\pi ^{2}\alpha \beta \,\delta (\Sigma _{12})\wedge \delta (\Sigma_{34})\,\mathbf{J}_{12}\;\mathbf{J}_{34} ~.
\end{equation}

However, it can be shown that the only solution of the Killing spinor equation with all additional  matter switched off is $\epsilon =0$, even locally. That means that adding a simple point-like matter as for codimension-two branes, whose purpose was to make local solution be valid globally, will not help in this case. The question whether adding some nontrivial gauge fields could stabilize this 0-brane remains open.

%%%%%%%%%%%%%%%%%%%%%%%%%%%%%%%%%%%%%%%%%%%%%%%%%%%%%%%%%%%%%%%%%%%%%%%%%%%%%%%%
\section{Codimension-two branes in CS supergravity  \label{Cod2}}
%%%%%%%%%%%%%%%%%%%%%%%%%%%%%%%%%%%%%%%%%%%%%%%%%%%%%%%%%%%%%%%%%%%%%%%%%%%%%%%%

%%%%%%%%%%%%%%%%%%%%%%%%%%%%%%%%%%%%%%%%%%%%%%%%%%%%%%%%%%%%%%%%%%%%%%%%%%%%%%%%
\subsection{Construction of a $(D-3)$-brane \label{killingDdim}}
%%%%%%%%%%%%%%%%%%%%%%%%%%%%%%%%%%%%%%%%%%%%%%%%%%%%%%%%%%%%%%%%%%%%%%%%%%%%%%%%

Consider the embedding space $\mathbb{R}^{D-1,2}$ with signature $(-,+,\cdots ,+,-)$ and global coordinates $x^{A}$ ($A=0,\ldots ,D$). The constraint $x\cdot x=-\ell ^{2}$ defines global AdS$_{D}$.

We introduce the following coordinate transformation, where the coordinates $x^{0},x^{1},x^{2},x^{D}$ resemble three-dimensional 0-brane $(B,\phi _{12},\phi _{0D})$, thickened by some extra coordinates $\rho_{u}$,
\begin{equation}
\begin{array}{l}
x^{0} = A\,\cos\phi_{0D} \cosh\rho_{1} \cdots \cosh\rho_{\frac{D-3}{2}} ~, \qquad x^{1} = B\,\cos\phi_{12} ~, \\ [1em]
x^{2} = B\,\sin\phi_{12} ~, \qquad x^{D} = A\,\sin\phi_{0D} \cosh\rho_{1} \cdots \cosh\rho_{\frac{D-3}{2}} ~,
\end{array}
\label{012D}
\end{equation}
where $B\in \lbrack 0,\infty )$, $\phi _{12}\in \lbrack 0,2\pi (1-\alpha ))$, $\phi _{0D}\in (-\infty ,\infty )$, noncompact coordinates $\rho_{u}\in \lbrack 0,\infty )$ ($u=1,\ldots ,\frac{D-3}{2}$) are radii of some cylinders, and
\begin{equation}
\begin{array}{l}
x^{2u+1} = A\cosh \rho _{1}\cdots \cosh \rho _{u-1}\sinh \rho _{u}\cos \phi_{2u+1,2u+2} ~, \\ [1em]
x^{2u+2} = A\cosh \rho _{1}\cdots \cosh \rho _{u-1}\sinh \rho _{u}\sin \phi _{2u+1,2u+2} ~,
\end{array}
\label{3...D-1}
\end{equation}
with azimuthal coordinates $\phi _{2u+1,2u+2}\in \lbrack 0,2\pi )$ ($u=1,\ldots ,\frac{D-3}{2}$) of these cylinders associated to the planes $x^{2u+1}$--$\,x^{2u+2}$.

Since the coordinate transformations (\ref{012D}, \ref{3...D-1}) preserve AdS constraint, the spacetime parameterized by $x^{\mu }=(B,\rho _{u},\phi _{ij})$ is locally AdS. However, because the angle $\phi_{12}$ in the 1--2-plane $\Sigma_{12}$ has an angular deficit $\alpha$, the global structure of the manifold has been changed, generating a conical singularity at $r=0$ that presents a ($D-3$)-brane.

The metric has the form
\begin{eqnarray*}
&& ds^{2} = \frac{\ell ^{2}}{A^{2}}\,dB^{2}+B^{2}d\phi_{12}^{2}-A^{2}\cosh^{2}\!\rho_{1} \cdots \cosh^{2}\!\rho_{\frac{D-3}{2}}\,d\phi_{0D}^2 \\ [1em]
&& \qquad \qquad + A^{2} \sum_{u=1}^{\frac{D-3}{2}} \cosh^{2}\!\rho_{1} \cdots \cosh^{2}\!\rho_{u-1} \left( d\rho_{u}^{2} + \sinh^{2}\!\rho_{u}\,d\phi_{2u+1,2u+2}^{2} \right) ~.
\end{eqnarray*}
The vielbein $e^{A}=e_{\mu }^{A}\,dx^{\mu}$ can be chosen as
\begin{equation}
\begin{array}{l}
e^{0} = A \cosh\rho_{1} \cdots \cosh\rho_{\frac{D-3}{2}}\,d\phi _{0D} ~, \qquad e^{1} = \dfrac{\ell}{A}\,dB ~, \qquad e^{2} = B\,d\phi_{12} ~, \\ [1.3em]
e^{2u+1} = A \cosh\rho_{1} \cdots \cosh\rho_{u-1}\,d\rho_{u} ~, \qquad e^{2u+2} = A \cosh\rho_{1} \cdots \cosh\rho_{u-1} \sinh\rho_{u}\,d\phi_{2u+1,2u+2} ~,
\end{array}
\label{e}
\end{equation}
and the torsionless spin connection out of the source then reads
\begin{equation}
\omega^{01} = \dfrac{B}{\ell A}\,e^{0} ~, \qquad \omega^{1,r+2} = - \dfrac{1}{\ell}\,e^{r+2} ~, \qquad \omega^{12} = - \dfrac{A}{\ell}\,d\phi_{12} ~, \qquad \omega^{2,r+2} = \omega^{0,2s} = 0 ~,
\end{equation}
and
\begin{equation}
\omega ^{0,2u+1}=\sinh \rho _{u}\cosh \rho _{u+1}\ldots \cosh \rho _{\frac{D-3}{2}}d\phi _{0D} ~,
\end{equation}
where $u=1,\ldots ,\frac{D-3}{2}$, $s=1,\ldots ,\frac{D-1}{2}$ and $r=1,\dots ,D-3$, and also
\begin{equation}
\begin{array}{l}
\omega^{2u+1,2v+2} = \omega^{2u+2,2v+2} = 0 ~, \qquad \omega^{2u+2,2u+1} = \cosh\rho_{u}\,d\phi_{2u+1,2u+2} ~, \\ [1.3em]
\omega^{2u+1,2v+1} = \sinh\rho_{v} \cosh\rho_{v+1} \cdots \cosh\rho_{u-1}\,d\rho_{u} ~, \\ [1.3em]
\omega^{2u+2,2v+1} = \sinh\rho_{v} \cosh\rho_{v+1} \cdots \cosh\rho_{u-1} \sinh\rho_{u}\,d\phi_{2u+1,2u+2} ~,
\end{array}
\label{w}
\end{equation}
with $u,v=1,\ldots ,\frac{D-3}{2}$ and $v<u$. The $so(2,D-1)$ curvature reads $\mathbf{F}=-\frac{A}{\ell }\,dd\phi _{12}\,\mathbf{J}_{12}$, and because of the deficit $\alpha $ in the range of the angle $\phi _{12}$, we have
\begin{equation}
\mathbf{F}=\mathbf{j}_{[D-3]}=2\pi \alpha \,\delta (\Sigma _{12})\,\mathbf{J}_{12} ~.
\end{equation}
%

%%%%%%%%%%%%%%%%%%%%%%%%%%%%%%%%%%%%%%%%%%%%%%%%%%%%%%%%%%%%%%%%%%%%%%%%%%%%%%%%
\subsection{Killing spinors without matter  \label{Dspinor}}
%%%%%%%%%%%%%%%%%%%%%%%%%%%%%%%%%%%%%%%%%%%%%%%%%%%%%%%%%%%%%%%%%%%%%%%%%%%%%%%%

In order to find a Killing spinor $\epsilon $ satisfying the equation $D \boldsymbol \epsilon =0$ whose gauge connection $\mathbf{A}$ is a $(D-3)$-brane, we use the known solution for $D=5$ for the coordinates $r=(1-\alpha )B$, $\rho _{1} $ and $\phi _{12}$. So we have in five dimensions
\begin{equation}
\epsilon =e^{-f(r)\Gamma _{1}}e^{-\frac{1}{2}\,\rho _{1}\Gamma _{3}}e^{\frac{1}{2}\,\phi_{12} \Gamma_{12}}\,\xi ~,
\end{equation}
with $f(r)$ given by Eq.(\ref{psi}).

In $D\geq 5$, we can proceed by mathematical induction. As we already know the solution with only one $\rho _{1}$ coordinate for the $D=5$ case, we assume the following dependence in odd $D$ for the first $u-1$ coordinates $\rho_{v}$,
\begin{equation}
\epsilon =e^{-f\Gamma _{1}}\prod_{v=1}^{1-u}e^{-\frac{1}{2}\,\rho_{v}\,\Gamma_{2v+1}}\xi_{u} ~,  \label{ind1}
\end{equation}
where $\xi_{u}$ is a spinor that depend on all variables but $\rho_{u}$. The $D$-dimensional $\Gamma $-matrices satisfy the Clifford algebra $\left\{ \Gamma_{a},\Gamma_{b}\right\} =2\eta _{ab}$.

Now the equation $D_{\rho _{u}}\epsilon =0$ takes the form
\begin{eqnarray*}
&& \left( \partial_{\rho_{u}} + \frac{1}{2\ell}\,\cosh\rho_{1} \cdots \cosh\rho_{u-1} (A - B\,\Gamma_1)\, \Gamma _{2u+1} \right. \\ [1em]
&& \qquad \qquad + \left. \frac{1}{2}\sum_{v=1}^{u-1}\sinh\rho_{v} \cosh\rho_{v+1} \cdots \cosh\rho_{u-1}\,\Gamma_{2u+1,2v+1}\right) \epsilon = 0 ~,
\end{eqnarray*}
where $\Gamma _{ab}:=\frac{1}{2}\,\left[ \Gamma _{a},\Gamma _{b}\right] $. Using Eq.(\ref{ind1}) and $e^{2f\Gamma_{1}}=(A+B\Gamma _{1})/\ell $ , the last expression is equivalent to
\begin{eqnarray*}
&& \left[ \partial_{\rho_{u}} + \frac{1}{2}\,\Gamma _{2u+1} \left( e^{\rho_{1}\Gamma _{3}} \cosh\rho_{2} \cdots \cosh\rho_{u-1} + \frac{1}{2} \sum_{v=2}^{u-1} \sinh\rho_{v} \cosh\rho_{v+1} \cdots \cosh\rho_{u-1}\,\Gamma_{2v+1} \right) \right] \\ [1em]
&& \qquad \qquad \qquad \qquad \qquad \times \prod_{v=1}^{1-u} e^{-\frac{1}{2}\,\rho_{v} \Gamma_{2v+1}}\,\xi_{u} = 0 ~.
\end{eqnarray*}
Multiplying $e^{\frac{1}{2}\,\rho_{1}\Gamma _{3}}$ by the left in the first term, it cancels the
$e^{-\frac{1}{2}\,\rho _{1}\Gamma _{3}}$ coming from the right. The same occurs for the last $u-2$ terms, since $[\Gamma _{3},\Gamma _{2u+1,2v+1}]=0$ and $v>1$. Furthermore, the $\rho _{1}$ dependence vanishes in the second term due to $e^{\frac{1}{2}\,\rho _{1}\Gamma_{3}} e^{\rho_{1}\Gamma _{3}}e^{-\frac{1}{2}\,\rho_{1}\Gamma_{3}}=1$, and thus the equation does not depend on the coordinate $\rho _{1}$,
\begin{eqnarray*}
&& \left[ \partial_{\rho_{u}} + \frac{1}{2}\,\Gamma_{2u+1} \left( e^{\rho_{2} \Gamma_{5}} \cosh\rho_{3} \cdots \cosh\rho_{u-1} + \frac{1}{2} \sum_{v=3}^{u-1} \sinh\rho_{v} \cosh\rho_{v+1} \cdots \cosh\rho_{u-1}\,\Gamma_{2v+1} \right) \right] \\ [1em]
&& \qquad \qquad \qquad \qquad \qquad \times \prod_{v=2}^{1-u} e^{-\frac{1}{2}\,\rho_{v} \Gamma_{2v+1}}\,\xi_{u} = 0 ~.
\end{eqnarray*}
We can now repeat the former procedure multiplying $e^{\frac{1}{2}\,\rho_{2}\Gamma_{5}}$ by left and the dependence on $\rho_{2}$ disappears, as well. In the same way all $\rho _{v}$ will vanish, leaving
\begin{equation}
\left( \partial _{\rho _{u}}+\frac{1}{2}\,\Gamma _{2u+1}\right) \xi _{u}=0 ~,
\end{equation}
whose solution is $\xi _{u}=e^{-\frac{1}{2}\,\rho _{u}\Gamma _{2u+1}}\xi _{u+1}$, and we recover for the next step in the inductive procedure the form (\ref{ind1}). This means that the spinor $\epsilon $ has the form
\begin{equation}
\epsilon =e^{-f(r)\Gamma _{1}}\prod_{u=1}^{(D-3)/2}e^{-\frac{1}{2}\,\rho _{u}\Gamma _{2u+1}}\varphi (\phi_{ij}) ~,
\end{equation}
where the spinor $\varphi $, that depends on all angular coordinates $\phi _{0D},\phi _{12},\ldots ,\phi_{D-2,D-1}$, has to be determined.

Let us solve now the component $D_{\phi _{2u+1,2u+2}}\epsilon =0$ of the Killing spinor equation,
\begin{eqnarray}
&& \left( \partial_{\phi_{2u+1,2u+2}}+\frac{1}{2}\,\cosh\rho_{u}\,\Gamma_{2u+2,2u+1} + \frac{1}{2\ell}\,\cosh\rho_{1} \cdots \cosh\rho_{u-1} \sinh\rho_{u} (A - B\,\Gamma_{1})\,\Gamma_{2u+2} \right. \nonumber \\ [0.9em]
&& \qquad \qquad + \left. \frac{1}{2} \sum_{v=1}^{u-1} \sinh\rho_{v} \cosh\rho_{v+1} \cdots \cosh\rho_{u-1} \sinh\rho_{u}\,\Gamma_{2u+2,2v+1} \right) \epsilon = 0 ~.
\end{eqnarray}
Doing exactly the same that was done to find out the $\rho _{u}$-dependence, we can get rid of the $\rho$ variables and finally we are left with
\begin{equation}
\left( \partial _{\phi _{2u+1,2u+2}}-\frac{1}{2}\,\Gamma _{2u+1,2u+2}\right) \varphi =0 ~, \quad u=1,\ldots, \tfrac{D-3}{2} ~,
\end{equation}
so that $\varphi \propto e^{\frac{1}{2}\,\phi _{2u+1,2u+2}\Gamma _{2u+1,2u+2}}$ and
$$
\epsilon = e^{-f\Gamma _{1}}e^{\frac{1}{2}\,\phi _{12}\Gamma _{12}}\prod_{u=1}^{\left( D-3\right)
/2}e^{-\frac{1}{2}\rho _{u}\Gamma_{2u+1}} \prod_{v=1}^{\left( D-3\right) /2}e^{\frac{1}{2}\,\phi _{2v+1,2v+2}\Gamma _{2v+1,2v+2}}\Phi (\phi_{0D}) ~,
$$
where we have restored the dependence on $\phi_{12}$ which can be added anywhere as long as it is placed after the $\Gamma_{1}$ exponential.

Finally, let us solve the equation $D_{\phi_{0D}}\epsilon = 0$,
$$
\left( \partial _{\phi _{0D}}+\frac{1}{2\ell }\Gamma _{0}\cosh \rho _{1}\ldots \cosh \rho
_{\frac{D-3}{2}}(A+B\Gamma _{1}) + \sum_{u=1}^{\frac{D-3}{2}}\sinh \rho _{u}\cosh \rho _{u+1}\ldots \cosh \rho_{\frac{D-3}{2}}\Gamma _{0,2u+1}\right) \epsilon = 0 ~.
$$
Proceeding as with the other equations, we obtain $\left( \partial_{\phi _{0D}} + \frac{1}{2}\,\Gamma _{0}\right) \Phi = 0$, leading to
\begin{equation}
\Phi = e^{-\frac{1}{2}\,\phi_{0D}\Gamma_{0}}\eta ~,
\end{equation}
where $\eta$ is a constant spinor. The final result is
$$
\epsilon_{s} = e^{-f(r)\Gamma _{1}}\prod_{u=1}^{(D-3)/2}e^{-\frac{1}{2}\,\rho_{u}\Gamma_{2u+1}}  \prod_{v=1}^{(D-3)/2}e^{\frac{1}{2}\,\phi_{2v+1,2v+2}\Gamma_{2v+1,2v+2}}\,e^{\frac{1}{2}\,\left(
\phi_{12}\Gamma_{12}-\phi _{0D}\Gamma_{0}\right)}\, \eta_{s} ~.
$$
Given the set of mutually commuting matrices $\Gamma_{2u+1,2u+2}$, with $(\Gamma_{2u+1,2u+2})^{2}=-1$, the constant spinors $\eta_{s}$ can be chosen as their common eigenvector,
\begin{equation}
\Gamma _{2u+1,2u+2}\,\eta _{s}=i\eta _{s} ~,\qquad \left( u=1,\ldots ,\tfrac{D-3}{3}\right) ~.
\end{equation}
The matrix $\Gamma_{0}$ is proportional to $\Gamma_{12}\,\Gamma_{34}\cdots \Gamma_{D-2,D-1}$ and it can always be normalized so that
\begin{equation}
\Gamma_{0}\,\eta_{s}=i\eta _{s} ~.
\end{equation}
The spinor $\epsilon_{s}$ then adopts the form
\begin{equation}
\epsilon _{s} = e^{-f(r)\Gamma _{1}}\prod_{u=1}^{(D-3)/2}e^{-\frac{1}{2}\,\rho _{u}\Gamma _{2u+1}} \prod_{v=1}^{(D-3)/2}e^{\frac{i}{2}\,\phi _{2v+1,2v+2}}\,e^{\frac{i}{2}\,(1-\alpha )\left( \phi -\frac{t}{\ell }\right)}\, \eta _{s} ~,
\end{equation}
where $\phi _{12}=(1-\alpha)\,\phi $ and $\phi _{0D}=(1-\alpha )t/\ell $. In order for spinor to be globally well-defined (periodic or antiperiodic in all angles), we need
\begin{equation}
\alpha =0 ~,
\end{equation}
which gives global AdS$_{D}$.

%\bibliographystyle{apsrev4-1}
%\bibliography{CS_Branes}

%merlin.mbs 2010-03-15 4.21a (PWD, AO, DPC)

\end{document}